\documentclass[showpacs, pra,twocolumn,preprintnumbers ,amsmath, amssymb, superscriptaddress, aps]{revtex4-2}
\usepackage{color}
\usepackage{amsmath,amssymb}
\usepackage{pifont}
\usepackage{amssymb}  
\usepackage{bbold}
\usepackage{float}
\usepackage{subfloat}

\usepackage[caption=false]{subfig}
\usepackage{tikz}
\usepackage{makecell}
\usepackage{subfig}
\usepackage{pifont}   
\usepackage{graphicx} 
\graphicspath{{Figures/}}
\usepackage{dcolumn}  
\usepackage{bm}       
\usepackage{multirow} 
\usepackage{placeins}
\usepackage[colorlinks]{hyperref}
\usepackage{mathtools}
\DeclarePairedDelimiter{\paren}{\lparen}{\rparen}
\newcommand{\vect}[1]{\mathbf{#1}}
\newcommand{\ket}[1]{\left|{#1}\right\rangle}
\newcommand{\bra}[1]{\left\langle{#1}\right|}

\newcommand{\eqfitpage}[1]{\resizebox{\linewidth}{!}{$#1$}}

\begin{document}
	
	\title{Chiral limits and effect of light on the Hofstadter butterfly in twisted bilayer graphene }
	\date{\today}
	\author{Nadia Benlakhouy}
	\affiliation{Laboratory of Theoretical Physics, Faculty of Sciences, Choua\"ib Doukkali University, PO Box 20, 24000 El Jadida, Morocco}
	\author{Ahmed Jellal}
	\affiliation{Laboratory of Theoretical Physics, Faculty of Sciences, Choua\"ib Doukkali University, PO Box 20, 24000 El Jadida, Morocco}
	\affiliation{Canadian Quantum  Research Center,
				204-3002 32 Ave Vernon,  BC V1T 2L7,  Canada}
	\author{Hocine Bahlouli}
		\affiliation{Physics Department, King Fahd University
		of Petroleum $\&$ Minerals,
		Dhahran 31261, Saudi Arabia}
		
			\author{Michael Vogl}
	\affiliation{Physics Department, King Fahd University
		of Petroleum $\&$ Minerals,
		Dhahran 31261, Saudi Arabia}
	\pacs{ 
	}
	\begin{abstract}
We study the magnetic field induced Hofstadter butterfly in twisted bilayer graphene (TBG) in various kinds of situations. First, we study the equilibrium case and identify the interlayer hopping processes that are most crucial for the appearance of a Hofstadter butterfly. Surprisingly, the hopping processes that are important for the appearance of the Hofstadter butterfly can be categorized as AA stacking type - that is interlayer hoppings between equivalent sublattices. This is in contrast to AB/BA-type hoppings that are important for the appearance of flat bands in magic angle TBG and were discussed in \cite{Tarnopolsky}. We also find that if AB-type interlayer-hopping processes are turned off the resulting model is chiral but differs from the model discussed in \cite{Tarnopolsky}. Therefore, TBG has two separate chiral limits - one of them is important to understand the formation of flat bands and the other for the Hofstadter butterfly. Taking this as motivation we discuss how the role of AA-type hoppings in combination with lattice relaxation effects can make individual Landau levels slightly harder to resolve in an experimental setting than one would expect from a non-relaxed lattice setting. Finally, we consider the impact of different forms of light on the fractal structure of the butterfly spectrum. Particularly, we study the impact of circularly polarized light and longitudinal light originating from a waveguide. As the system is exposed to circularly polarized light we find butterflies with increasingly pronounced asymmetry with respect to energy $E=0$. This is due to the introduction of a gap term that breaks the chiral symmetries for both of the two chiral limits mentioned above. Lastly, we study the effect of longitudinal light that can be produced at the exit of a waveguide, in a slightly simplified model. Here, we find that no additional terms that break chiral symmetry are introduced. Therefore, it is found to lead to no increase in asymmetry of the energy spectrum. In fact, we identify specific experimentally accessible driving regimes in which the TBG achieves any of the two chiral limits.
	\end{abstract}

	\maketitle
\section{Introduction}
	
The developments that led to our present work go all the way back to graphene, a material that was first synthesized when it was peeled off a graphite substrate using a scotch tape technique \cite{Novoselov}. It can be viewed as a monolayer of carbon atoms arranged in a honeycomb crystal lattice structure. This seemingly simple structure is at the origin of marvelous electronic and optical properties that made scientist predict that graphene might revolutionize the nanotechnology industry due to its  potential for the development of more efficient electronic components \cite{CastroNeto}. Indeed, one of graphene's most exciting and counter intuitive transport properties is that is it allows for Klein tunneling: charges can tunnel through an electrostatic barrier regardless of its height. This effect while it is interesting from a perspective of power consumption has a huge drawback because it means that electrons are difficult to confine. Hence, it thus far - in the case of graphene - prohibits the realization of switching devices such as field effect transistors (FET) which are the basic building blocks of modern electronics. Therefore, a lot of efforts have been devoted to overcome this difficulty via band engineering \cite{Wang, San, Kindermann, Zarenia, Zhou, Costa}. One of the simplest approaches to modify graphene's band structure is by stacking multiple graphene layers on top of one another. The simplest case is different stacking configurations of bilayer graphene (BLG) which consists of two stacked single layers. Bilayer graphenes have continued to attract significant theoretical interest in recent years \cite{Snyman, Barbier, Ben1, Ben, MacCann, Rozhkov, Ohta, Benlakhouy, Chiu, Zahidi,Naumis2021,Navarro1,Navarro2}.

Recent technological advances have made it possible to tune the electronic properties of layered materials without changing the atomic structure of the individual layers. The simplest technique in this regard is twisting successive layers with respect to one another. Such a twist creates an angle-dependent moir\'{e} pattern, a periodic pattern of relatively giant unit cells -  a size $\sim 1000$ times the size of AB bilayer graphene's unit cells is not atypical. Moreover, the twist angle plays the role of a knob that affects the carrier interaction. In fact, it has been discovered that at certain, so-called magic angles, flat energy bands occur which then give rise to a plethora of highly correlated phases. An exciting example is a superconducting phase that occurs at the first magic twist angle \cite{Cao1,Cao, Shen, Liu, Lee, Culchac, Kerelsky, Rubio, Halbertal, Christos, Wong, Lu, Sharpe, Seo}. This superconducting state was first discovered in a land-mark paper in Nature \cite{Cao}. We stress that this observation can be traced back to the appearance of moir\'{e} patterns, which are the source of flat band behavior that eventually gives rise to superconductivity in twisted bilayer graphene (TBLG). It should be mentioned that it was found  early on - using scanning tunneling microscope (STM) techniques \cite{Hass, Brihuega, Meng, Luican1, Yan, Li1, Cisternas} - that these moir\'{e}-type patterns can exist in multilayer graphene. Thus, twisted bilayer graphene not only offers the possibility to engineer the electronic band structure but also allows to control the relative importance of carrier correlations. 

Last, but not least, band structures can also be modified by the application of periodically time-dependent external drives \cite{Seo, Oka, Oka1, Luo, Kibis}. In fact, for the case of graphene it has been shown theoretically that the drastic effect of strong high-frequency electromagnetic fields on the electron energy spectrum near the Dirac point depends strongly on polarization of the field. Linear polarization results in an anisotropic gapless energy spectrum while circular polarization gives rise to an isotropic energy gap. Hence, the transport properties of electrons also in twisted bilayer graphene are strongly affected by the polarization of the electromagnetic waves. From the application point of view it is easier to control light intensity and polarization of an external light source than the twisting angle. Therefore, light can be seen as an ideal control knob for the transport properties of twisted bilayer systems.

Floquet theory has been widely adapted to treat the electromagnetic interaction between external light sources and materials and make these types of predictions. In this context various techniques have been developed to achieve a time-independent description \cite{Vogl4, Albanin, Itin, Mikami, Mohan, Vogl5, Vogl1, Vogl6, Vega, Vega1, Kennes, Li11, Muller}. More recently, the moir\'{e} and Floquet approaches have been combined to make theoretical predictions about various topological phases in moir\'{e} materials \cite{Ibsal, Topp, Vogl2, Vega2, Lu1, Li1}. 

Important for our current investigation is the recent work of some of the current authors who studied TBLG under the influence
of two distinct forms of light polarization, circularly polarized light and transverse magnetic (TM) light emanating from a waveguide  \cite{Vogl1,Vogl2}. For the circularly polarized light case a rotating frame Hamiltonian, valid for both weak and strong drives in the high to intermediate drive frequency regime, has been developed \cite{Vogl1}. This Hamiltonian is appropriate in the regimes where the ordinary Van Vleck approximation ($\text{vV}$) breaks down, and results in a significantly enhanced approximation to the quasienergies \cite{Vogl4}.
 
In the present work we wish to build on these developments and apply the effective Hamiltonian approach \cite{Vogl1,Vogl2} to another interesting phenomenon that have been investigated in the context of twisted bilayer graphene since the start of this field \cite{MacDonlad} and has recently begun to rise to popularity again \cite{Arbeitman1,Arbeitman2,Kasra,andrews_Fractional_2020}. In particular, we consider the so-called Hofstadter butterfly phenomenon \cite{Hofstadter}, where electrons under the influence of a magnetic field exhibit an energy spectrum that displays fractal patterns. These types of fractal butterfly spectra have been observed experimentally in monolayer graphene (MLG) \cite{Hunt, Ponomarenko}, in AB-BLG, placed on a hexagonal boron nitride (hBN) substrate \cite{Dean}, also on square, honeycomb and triangular lattices \cite{Oh, Oh1}, kagome lattices \cite{Du} and in TBLG \cite{Bistritzer, Kasra, Zhang1,Arbeitman2}. While the effect of a periodic drive - in our case light - on the Hofstadter butterfly has been studied in various systems \cite{Wang, Wang2, Lawton, Wang3, Labadidi, Zhou1, Ding, Wackerl, Kooi}. To our knowledge the Hofstadter butterfly in TBLG has not been extensively studied. Specifically, previous works have mostly focused on the kicked-Harper model \cite{Wang, Wang2, Lawton, Wang3}. Other cases include MLG subjected to a uniform perpendicular magnetic field $B$ in combination with a laser. Here, the Floquet Hofstadter butterfly exhibits a richer structure than in the equilibrium case \cite{Ding, Wackerl, Kooi}. While not extensively studied in TBLG there have been some studies that investigated the Hofstadter butterfly under the influence of light\cite{Pilkyung1,Pilkyung2}. While these works studied extensively the fractal properties of Landau levels under the influence of light, here we want to take a slightly different route and study the interplay between light, chiral symmetries and the fractal properties.

The work is organized as follows. In Sec. \ref{Model}, we describe TBLG, introduce the theoretical model, and highlight some of its equilibrium properties. In Sec. \ref{sec:analysis_eq_param}, we analyse which of TBLG's hopping processes is most important for the appearance of a Hofstadter butterfly at low energies and how the different hopping processes influence the symmetry with respect to energy $E=0$. By inspection of the symmetry properties in twisted bilayer graphene we find that it has two separate chiral limits. In
Sec. \ref{Circularly polarized light}, we describe the first form of light - circularly polarized light - which is used in the Hamiltonian description, and  discuss our numerical results. In Sec \ref{Waveguide light}, we consider longitudinal light emanating from a waveguide and discuss our numerical results. In Sec. \ref{Summary and conclusion}, we summarize our main results and present our conclusions.

	\section{Equilibrium Case}
	%
\label{MODEL HAMILTONIANS}
\subsection{Model}\label{Model}
\begin{figure}[ht]
	\centering
	\subfloat[]{\includegraphics[width=0.46\linewidth]{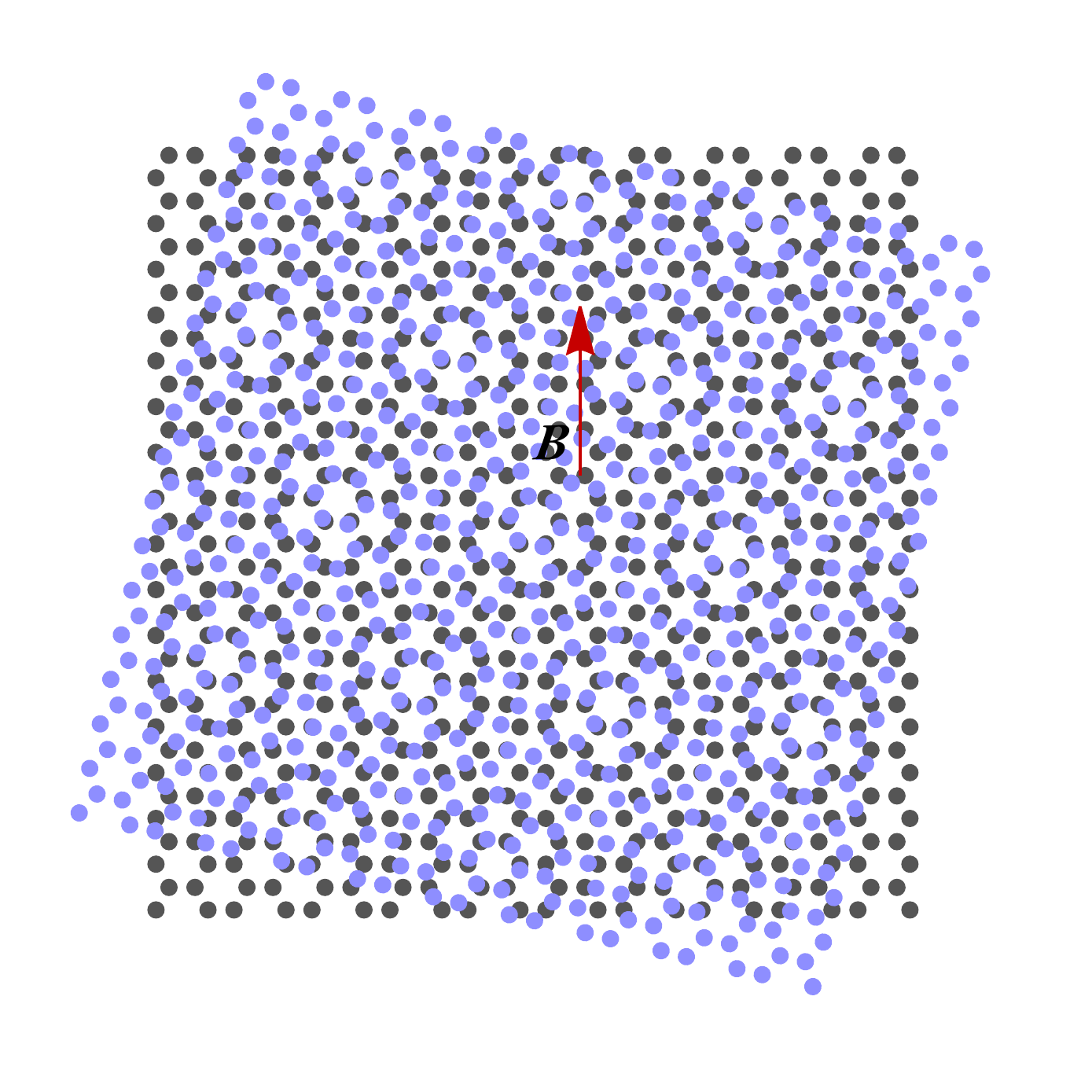} \label{fig:TBG_combined_figa}}
	\subfloat[]{\includegraphics[width=0.49\linewidth]{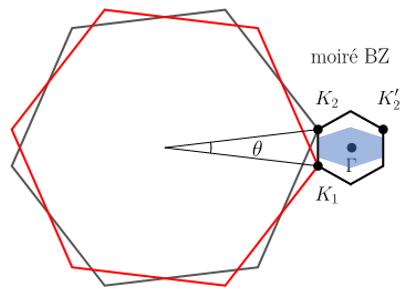}\label{fig:TBG_combined_figb}}\\
	\subfloat[]{\includegraphics[width=\linewidth]{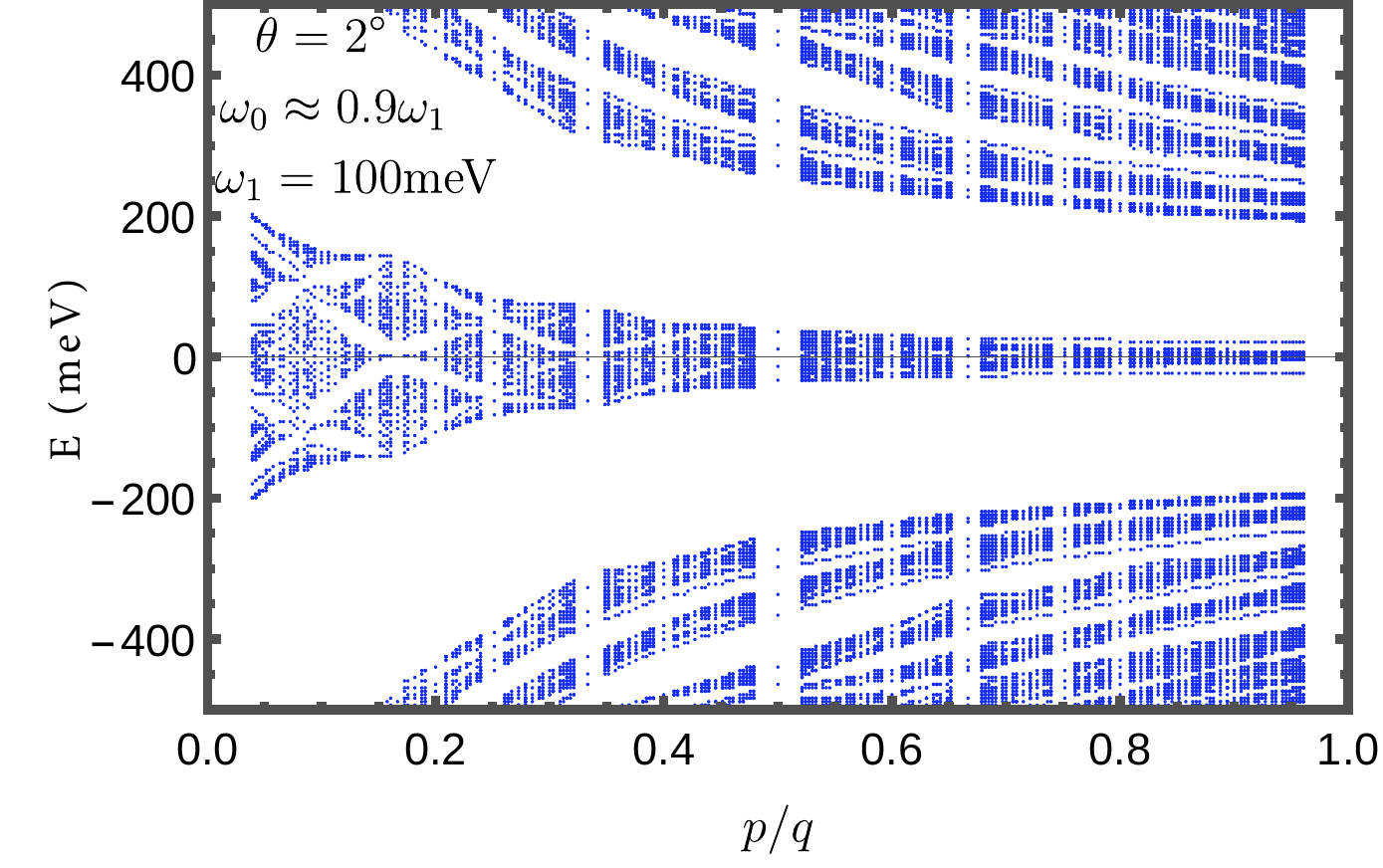}\label{fig:TBG_combined_figc}}
	\caption{(Color online) (a) A sketch of twisted bilayer graphene (TBLG). (b) The moir\'{e} Brillouin zone and a schematic depiction of the corresponding magnetic Brillouin zone in blue, which was obtained by reshaping the moir\'{e} Brillouin zone in Eq. \ref{eq:Brillouin_mag_zone}.
	(c) The Hofstadter butterfly in TBLG. }\label{fig:TBG_combined_fig}
\end{figure}

We start by considering the simplest case where two graphene layers are miss-aligned with respect to each other by an angle $\theta$. For this work we will work with a model that is valid for $\theta\precsim 10^{\circ}$ and will focus on the case of $\theta=2^\circ$, which lies well within its range of validity. The structure of the lattice we consider can be seen in Fig. \ref{fig:TBG_combined_figa}.

Following the approach used in Refs. \cite{Bistritzer, MacDonlad, Wu, Rost, Fleischmann, Xie} the Hamiltonian describing TBLG is given by
\begin{equation}
H(\mathbf{x})=\begin{pmatrix}
h(-\theta/2) & T(\mathbf{x}) \\
T^{\dagger}(\mathbf{x}) & h(\theta/2)
	\end{pmatrix},
\end{equation}	
where the intralayer Hamiltonian is given by
\begin{equation}
	h(\theta)=v_F\boldsymbol{p}\cdot\boldsymbol{\sigma}_{\theta}
\end{equation}
and rotated Pauli matrices are defined by
\begin{align}
 \boldsymbol{\sigma}_{\theta}=\paren[\bigg]{\cos\theta\sigma_x-\sin\theta\sigma_y,  \cos\theta\sigma_x+\sin\theta\sigma_y}.
\end{align}
The interlayer Hamiltonian describing  the tunneling processes between layers is given by  
\begin{equation}
	T(\mathbf{x})=\sum_{j=1}^{3} T_{j} \exp \left(-i \boldsymbol{q}_{j} \cdot \mathbf{x}\right),
\end{equation}
where  $\boldsymbol{q}_{1}=k_{\theta}(0,-1)$,  $\boldsymbol{q}_{2}=k_{\theta}(\sqrt{3},1)/2$,  $\boldsymbol{q}_{3}=k_{\theta}(-\sqrt{3},1)/2$ are the nearest neighbor vectors of the moir\'{e} Brillouin zone. The characteristic momentum-scale in the moir\'{e} Brillouin-zone is set by $k_{\theta}=2k_D\sin(\theta/2)$ with $k_D=4\pi/3a_0$ being the Dirac momentum, and $a_0=2.46  \textup{~\AA}$ is the lattice constant. The hopping matrices $T_i$ can be expressed in terms of Pauli matrices 
\begin{align}	
	&T_1=w_0\sigma^0+w_1\sigma^x,\\
	& T_2=w_0\zeta^*\sigma^0+w_1\sigma^{+}+w_1\zeta\sigma^-, \\
	& T_3=w_0\zeta\sigma^0+w_1\sigma^{+}+w_1\zeta^*\sigma^-,
\end{align}
where $\zeta=e^{2\pi i/3}$ and $\sigma^{\pm}=(\sigma^x\pm i\sigma^y)/2$ act on sublattice degrees of freedom. The $2 \times 2$ Pauli matrices and
identity matrix are denoted by $\sigma^{x,y,z}$ and $\sigma^0$, respectively.

We stress that the Hamiltonian introduced here is valid for the moir\'{e} BZ that is located near the $K$ point. The Hamiltonian describing the bands near the $K^\prime$ point is related by time reversal symmetry or equivalently it can be obtained by the replacements \begin{equation}
\begin{aligned}
&\vect q_i\to-\vect q_i;\quad
\zeta\to \zeta^*\\
&\boldsymbol{\sigma}_{\theta}\to\paren[\bigg]{-\cos\theta\sigma_x-\sin\theta\sigma_y,  -\cos\theta\sigma_x+\sin\theta\sigma_y}
\end{aligned},
\end{equation}
which was also noted in Appendix A of \cite{Kasra2}. We note that for the main part of the paper we will focus on the $K$ point to keep our discussion more coherent. Although, we found it instructive to provide results for the $K^\prime$ point in appendix \ref{app:kprime}.

A few comments are in order about the parameters $w_i$. Some stacking configurations in TBLG are more energetically favorable than others. Most importantly $\text{AB}$ and $\text{BA}$ stacking is energetically preferred compared to $\text{AA}$ stacked regions \cite{Fleischmann1, Nam}. This leads to different sizes of AA and AB stacked regions - at very small twist angles AB stacked regions grow in size compared to AA stacked regions \cite{Fleischmann1, Nam}. Furthermore, the different stacking regions have differing interlayer lattice spacings. We account for these effects in an approximate fashion through the parameters $w_i$ in the interlayer tunneling terms \cite{Vogl2, Li}.  In our specific case we take the hopping amplitudes as $w_0\approx 90$ meV for $\text{AA}$-type hoppings and  $w_1\approx 100$ meV for $\text{AB}/\text{BA}$-type hoppings. 

Next, let us discuss this Hamiltonian in the presence of a perpendicular magnetic field $\boldsymbol{B}=B_z\hat z$. This type of field is introduced through the minimal substitution procedure $\boldsymbol{\hat{p}}\longrightarrow\boldsymbol{\hat{p}}+e\boldsymbol{A}$. For computational convenience we select the Landau gauge $\boldsymbol{A}=B(-y,0,0)$. In this case we find that the intra-layer blocks of our Hamiltonian can be written in terms of ladder operators defined by
\begin{equation}
a=\frac{\ell}{\sqrt{2}}\left[p_x-eBy-ip_y  \right],\quad
a^\dagger=\frac{\ell}{\sqrt{2}}\left[p_x-eBy+ip_y  \right],
\end{equation}
which fulfill the usual commutation relation $\left[a, a^{\dagger}\right]=1$.
In this case the intralayer Hamiltonian becomes
\begin{equation}
	h(\theta/2)=\omega_{c}\left[\sigma^{+}e^{i\theta/2}a+\sigma^{-} e^{-i\theta/2}a^{\dagger}\right],
\end{equation}
where the $ \omega_c=\sqrt{2}v_F/\ell$ is the cyclotron energy, and  $ \ell=1/\sqrt{eB}$ is the magnetic length.
 
Since the dominant energy scale is given by the intralayer contributions it is now convenient to express the full Hamiltonian in a basis of layer index  $L=1, 2$, sublattice index $\alpha= A, B$, guiding center $y$ and Landau level $n$ degrees of freedom using a ket  $\ket{ L,n,\alpha, y}$. This ket, however, will not turn out to be the most convenient basis choice because interlayer terms can shift the guiding center by $\pm\Delta$ with $\Delta=\sqrt{3}k_\theta \ell^2/2$ - specifically through processes associated with $T_{2,3}$. This can easily be seen if we compute matrix elements 
\begin{equation}
    \bra{A/B,n,\alpha, y}e^{-i\vect q_{2,3}\vect x}\ket{B/A,n,\alpha, y^\prime}\propto \delta_{y,y^\prime\pm\Delta}.
    \label{eq:rel_element_for_periodicity}
\end{equation} 

By analogy to a tight binding model we can now directly see that the system has a periodicity arising from this restriction to interlayer hopping processes - a magnetic unit cell arises. Now for the resulting Hamiltonian to be diagonalizable for an infinite size system it is important that the moir\'{e} unit-cell is commensurate with the magnetic unit-cell because otherwise one will be dealing with a quasi-periodic system. It is found to be the case when the magnetic flux $\Phi$ through a unit-cell is such that \cite{MacDonlad}
\begin{equation}
	\Phi=\frac{q}{p} \phi_{0}, \quad \phi_{0}=\frac{h c}{e},
\end{equation}
where $p/q \in \mathbb{Q}$ is the rational number relating the flux $\Phi$ to the flux quantum $\phi_0$, in our case we have chosen units such that $\hbar=c=1$. It is found that the resulting magnetic moir\'{e} Brillouin zone (MMBZ) is bounded by
\begin{equation}
	0<k_{x}=\frac{ y_{0}}{ \ell^{2}}<\frac{\Delta}{\ell^2} , 
	\quad 0< k_y<\frac{2 \pi}{q\Delta},
\end{equation}
or rewritten to allow a comparison to the non-magnetic case
\begin{equation}
	0<k_{x}=\frac{ y_{0}}{ \ell^{2}}<\frac{\sqrt{3}}{2}k_\theta,
	\quad 0< k_y<\frac{k_\theta}{2p}.
	\label{eq:Brillouin_mag_zone}
\end{equation}

Making use of the periodicity seen in Eq. \eqref{eq:rel_element_for_periodicity} it is convenient to express the guiding center coordinate as $y=y_0+(mq+j)\Delta$, where $j\in{0,1,\dots,q-1}$. A computationally convenient basis is then given by the Fourier transform
\begin{equation}
    \ket{L,n,\alpha,j}=\frac{1}{\sqrt{N}}\sum_m e^{ik_y (mq+j)\Delta}\ket{L,n,y_0+(mq+j)\Delta},
    \label{eq:magn_basis}
\end{equation}
where we dropped the transformed label $k_y$ from the ket on the left side of the equation to simplify the notation because the Hamiltonian will be diagonal in $k_y$.
The intralayer Hamiltonian in this basis is then given as 
\begin{equation}
		\eqfitpage{h(\theta/2)=-\omega_{c} \sum\limits_{L, n, j}\left(e^{-i\theta_L/2} \sqrt{n+1}\ket{L, n+1, A, j}\bra{L,n, B, j}+\mathrm{H.c.}\right)}
		\label{eq:htheta}
\end{equation}
The interlayer Hamiltonian in the same basis can be expressed as 
\small	\begin{align}
	&T(\mathbf{k})=\sum_{n'n \alpha \beta j}\left[T_{1} F_{n' n}\left(\mathbf{z}_1\right) e^{-i k_x k_{\theta} \ell^2} e^{-4 \pi i \frac{p}{q} j}\ket{2 n' \alpha j}\bra{ 1 n \beta j}\right. \nonumber \\ 
	&+T_{2}F_{n' n}\left(\mathbf{z}_{2}\right) e^{i k_{y} \Delta} e^{\frac{i}{2} k_x k_{\theta} \ell^2} e^{i \pi \frac{p}{q}(2 j-1)}\ket{ 2 n' \alpha, j+1}\bra{ 1 n \beta j} \nonumber \\
	&\left.+T_{3}F_{n' n}\left(\mathbf{z}_{3}\right) e^{-i k_{y} \Delta} e^{\frac{i}{2} k_x k_{\theta} \ell^2} e^{i \pi \frac{p}{q}(2 j+1)}\ket{ 2 n' \alpha j-1}\bra{ 1 n \beta j}\right]
	\label{eq:Thopmagbas}
	\end{align}	
where  $ \mathbf{z}_{j}=\frac{q_{j x}+i q_{j y}}{\sqrt{2}}\ell$, and
\begin{equation}
\begin{aligned}
&F_{nm}(\vect z)=\begin{cases} \tilde F_{nm}(\vect z)& n\geq m\\
\tilde F_{nm}^*(-\vect z)& m<n
\end{cases}\\
&\tilde F_{nm}(\vect z)=\sqrt{\frac{m!}{n!}} e^ {-\frac{\vect z^2}{2}} (-z_1+i z_2)^{n-m} \mathcal{L}_m^{n-m}\left(\vect z^2\right)
\end{aligned},
\end{equation}
where $\mathcal{L}$ is the so-called associated Laguerre polynomial and $z_i$ the components of $\vect z$.

It is important to expand on one subtlety about the numerical implementation of this Hamiltonian that was mentioned as a brief footnote in \cite{MacDonlad}. While the Hamiltonian for the most part is straightforward to implement numerically, one has to be careful about the inclusion of basis states to avoid a spurious degeneracy at low energies. Particularly, let us  consider the case where we neglect interlayer couplings. Since we have chosen a model valid near the K point of graphene we have to realize that there is only one zero energy eigenstate per layer and K point. However, if we naively choose our basis states from
\begin{equation}
    \{L\in\{t,b\},\alpha\in\{A,B\},n\in\{0,\dots,n_{\mathrm{max}}\}\},
\end{equation}
and diagonalize numerically, we find additional spurious states at zero energy. To understand this better we may now recall the analytical expressions for the wavefunctions of the zero energy Landau level for graphene at the $K$ point. For each layer we find $\ket{n,\pm}=(\pm \ket{n-1},\ket{n})$. That is, $n=0$ only has contributions from sublattice $B$. Now, we find that there is zero energy states that have contributions from sublattice $A$, which we can identify as a numerical artifact coming from our choice of basis, which does not break sublattice symmetry. Clearly sublattice symmetry is broken by the solutions and we have to ensure this is enforced in our numerical approach. The way to achieve this is to make a slightly altered choice of basis states that explicitly breaks sublattice symmetry. This choice is given below
\begin{equation}
    \{L\in\{t,b\},\alpha\in\{A,B\},n\in\{0,\dots,n_{\mathrm{max}}-\delta_{\alpha,B}\}\},
\end{equation}

where the term $\delta_{\alpha,B}=1$ if the sublattice index $\alpha$ corresponds to sublattice $B$. That is sublattice $B$ has a smaller Landau level cut-off than sublattice $A$.

This explicit breaking of sublattice symmetry in the choice of basis states shifts spurious states to high energies, which will be of no consequence for us \cite{MacDonlad}. We stress that while, in the case of non-coupled layers the spurious low lying levels do not seem to matter in a plot of Landau levels (since the plot does not show degeneracy), this point becomes very important in the presence of  interlayer coupling. Indeed, as interlayer couplings get introduced Landau levels split and spurious low energy bands have a devastating effect. Therefore, it is of utmost importance to remove the spurious contributions using the approach we have just outlined.

\subsection{Equilibrium properties: Interlayer hoppings and the Hofstadter butterfly}
\label{sec:analysis_eq_param}
Next, we want to study some of the equilibrium properties of this model. First, we recall that it has been noted in \cite{MacDonlad} that in the presence of a non-zero magnetic field we find that the Landau levels in TBLG possess a fractal, self-similar structure see Fig. \ref{fig:TBG_combined_figc}.

\begin{figure}[htb!]
	\centering
	\subfloat[]{\includegraphics[width=0.8\linewidth]{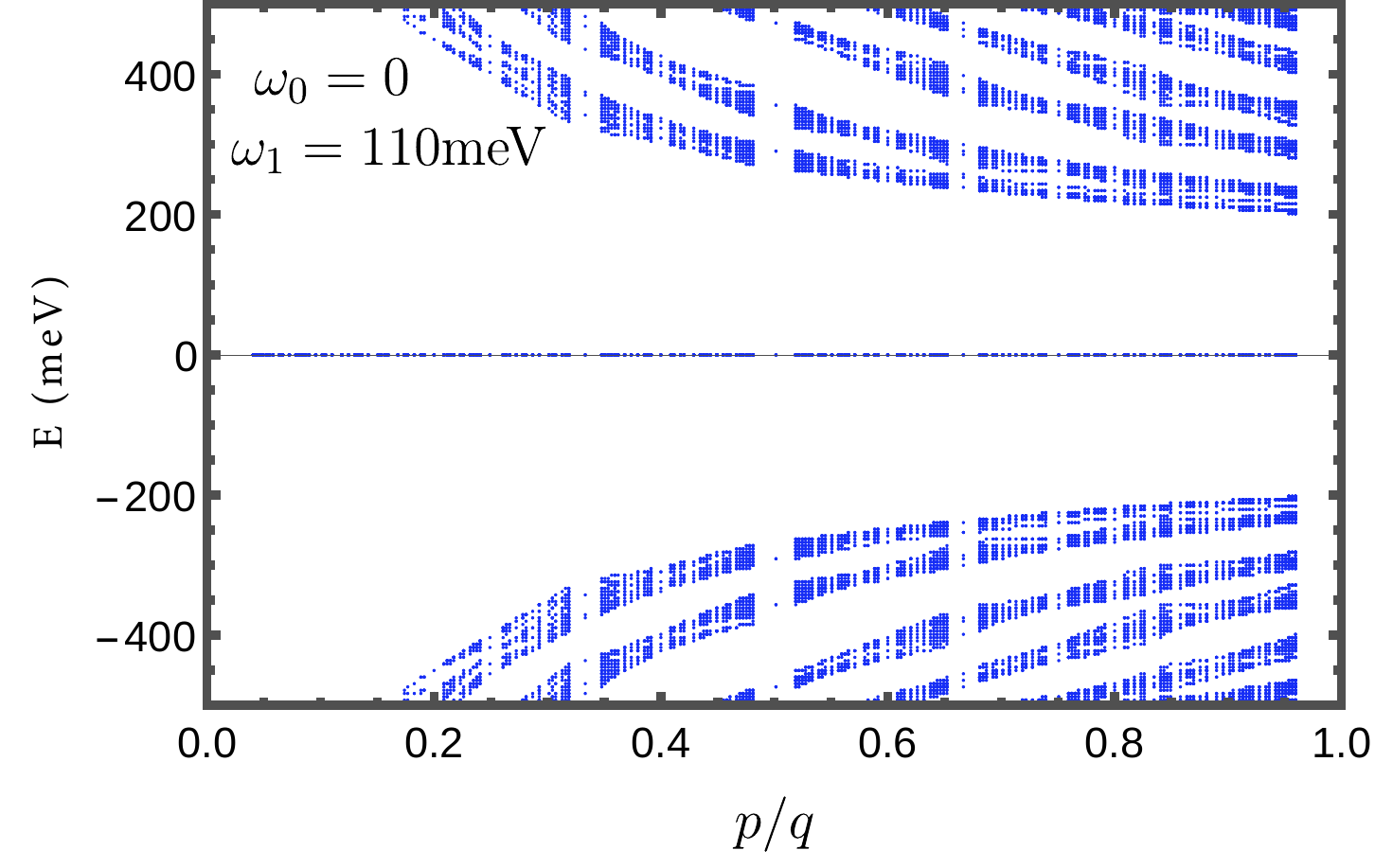}\label{fig:Hofstadter_BF__Func_InterlayerHoppinga}}\\
	\subfloat[]{\includegraphics[width=0.8\linewidth]{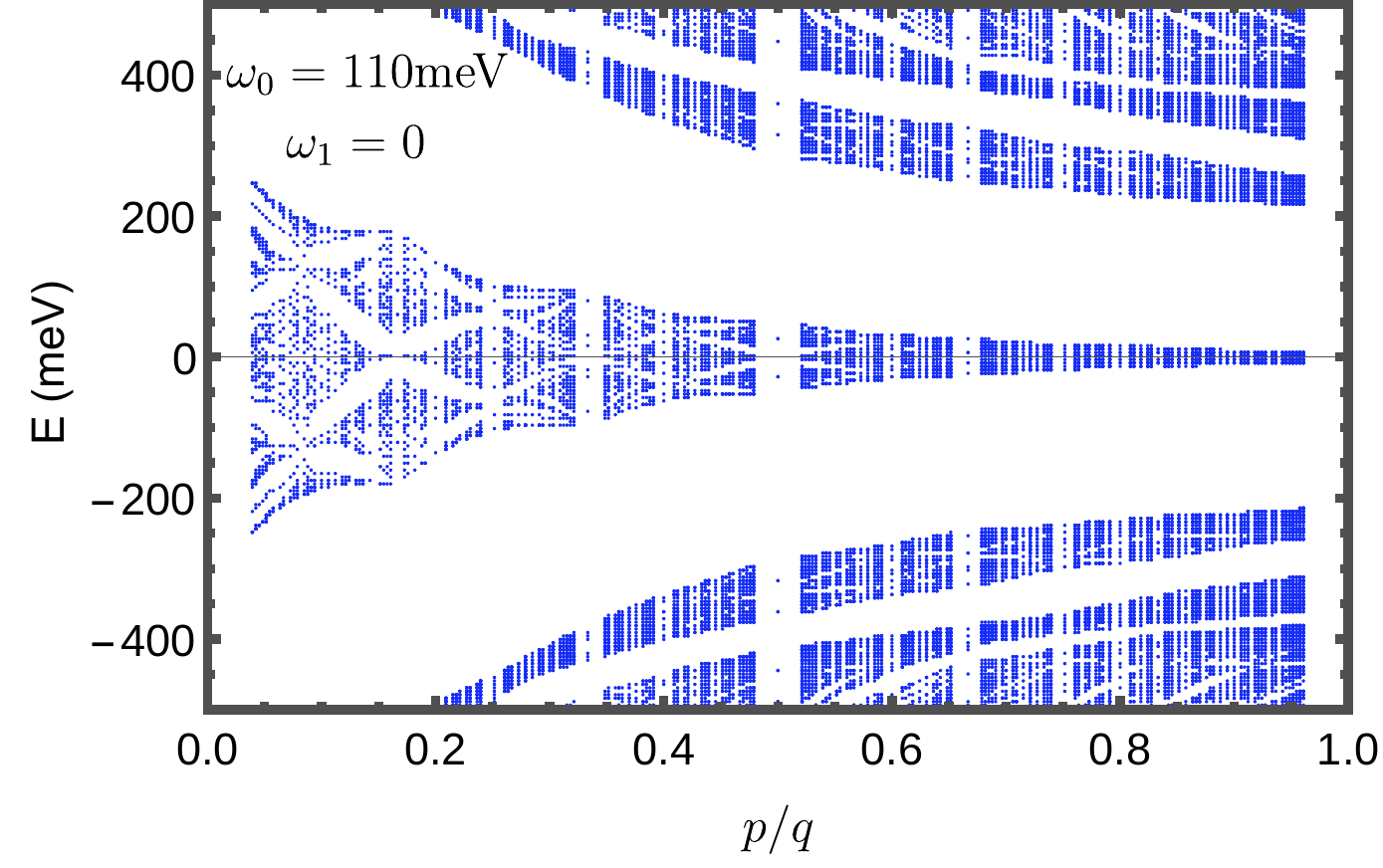}\label{fig:Hofstadter_BF__Func_InterlayerHoppingb}}
	\caption{{(Color online) The Floquet Hofstadter butterfly spectrum as function of $p/q=\phi_0/\Phi$ in the equilibrium case for different values of the hopping amplitude. (a) $w_0=0$, $w_1=110$ meV. (b) $w_0=110$ meV, $w_1=0$ meV.} A cutoff for the Landau level index $n$ that was used for the plots is the same as the one suggested in \cite{ MacDonlad} $n_{\mathrm{max}} \approx 2\left[\max \left(a_{0}\gamma_{\text{RF}}, w_1\right) / \omega_{c}\right]^{2}$.}\label{fig:Hofstadter_BF__Func_InterlayerHopping}
\end{figure}

We next want to determine which types of interlayer hopping processes are the most important for the appearance of the Hofstadter butterfly. Therefore, in Fig. \ref{fig:Hofstadter_BF__Func_InterlayerHopping} we plot the Hofstadter butterfly for different values of the $\text{AA}$-type hopping amplitudes $w_0$ and $\text{AB/BA}$-type hopping amplitudes $w_1$.

In Fig. \ref{fig:Hofstadter_BF__Func_InterlayerHoppinga} we observe that if we set $w_0=0$ meV, $w_1=110$ meV, which corresponds to the so-called chiral model \cite{Tarnopolsky}. For this case the Hofstadter butterfly at the center of our plot collapses into a zero energy line. The reason that the lowest Landau level collapses can be understood from a perturbative perspective. In essence is due to the fact that the lowest Landau level of graphene lives only on a single sublattice and that the terms proportional to $w_1$ in $T(\vect r)$ couple between sublattices. Therefore, these terms have no effect on the lowest Landau level, which is in stark contrast to other levels, which live on both sublattices. More precisely, treating  $V=T(\vect r)$ as a perturbation, the eigen-bi-spinors of the lowest Landau level for $T(\vect r)=0$ are $\ket{L_{01}}=(0,\ket{0},0,0)$ or $\ket{L_{02}}=(0,0,0,\ket{0})$. The first correction to the energy because of this can be found as $\bra{L_{02}}V\ket{L_{02}}=0$ - of course for $w_0=0$. Therefore, the lowest Landau level is left unsplit.

To contrast the case $w_0=0$ meV we also studied the opposite case of $w_0=110$ meV, $w_1=0$ in Fig. \ref{fig:Hofstadter_BF__Func_InterlayerHoppingb}. Interestingly, we find that in our case it is this term that leads to the splitting of the central Landau level and the appearance of the Hofstadter butterfly.  This is in stark contrast to the physics that leads to the appearance of flat bands, where this term is the less important one \cite{Tarnopolsky}.

We should also note that both reduced models $w_0=0$ or $w_1=0$ lead to a Hofstadter butterfly that is symmetric with respect to the axis $E=0$. This is because in these cases - unlike the case where both $w_i\neq 0$ - there exist unitary operators $C_i$ with $C_i^2=1$ that anticommute with the Hamiltonian. In the case $w_0=0$ it is $C_1=\mathrm{diag}(1,-1,1,-1)$ and in the case of $w_1=0$ it is $C_2=\mathrm{diag}(1,-1,-1,1)$. Indeed, from $C_iH\psi_n=-HC_i\psi_n$ one can directly see that each state $\psi_n$ fulfilling $E_n\psi_n=H\psi_n$ has a chiral partner state $C_i\psi_n$ with energy $-E_n$ and hence the spectrum is symmetric \cite{Yang2020}. 

Now the disappearance of the center butterfly in the case $w_0=0$ meV, $w_1=110$ meV leads to an intuitive understanding of some subtle possible experimental consequences for the measurement of an equilibrium Hofstadter butterfly in a relaxed TBG lattice. Particularly, we may realize that $w_0$ tunes the strength of the lowest Landau level splitting. This is something that becomes important to recognize because it has been found that $w_0$-type hoppings become less important as one reduces the twist angle in TBG \cite{carr}. This effect, which is due to lattice relaxations, can be captured by a fit of $w_0/w_1$ for angles $\theta=0.18$ to $6^\circ$ to the data in Fig. 3c of \cite{carr} as
\begin{equation}
    w_0/w_1\approx\frac{\tan ^{-1}\left(\frac{0.0236}{\theta }\right)}{-0.0002\, +\frac{0.00001}{\theta ^2}+\frac{0.024}{\theta }}.
\end{equation}
From here we can see that $w_0$ shrinks at small twist angles like it is shown in the Fig. \ref{fig:w0overw1} below.

\begin{figure}
	\begin{center}
	\includegraphics[width=0.8\linewidth]{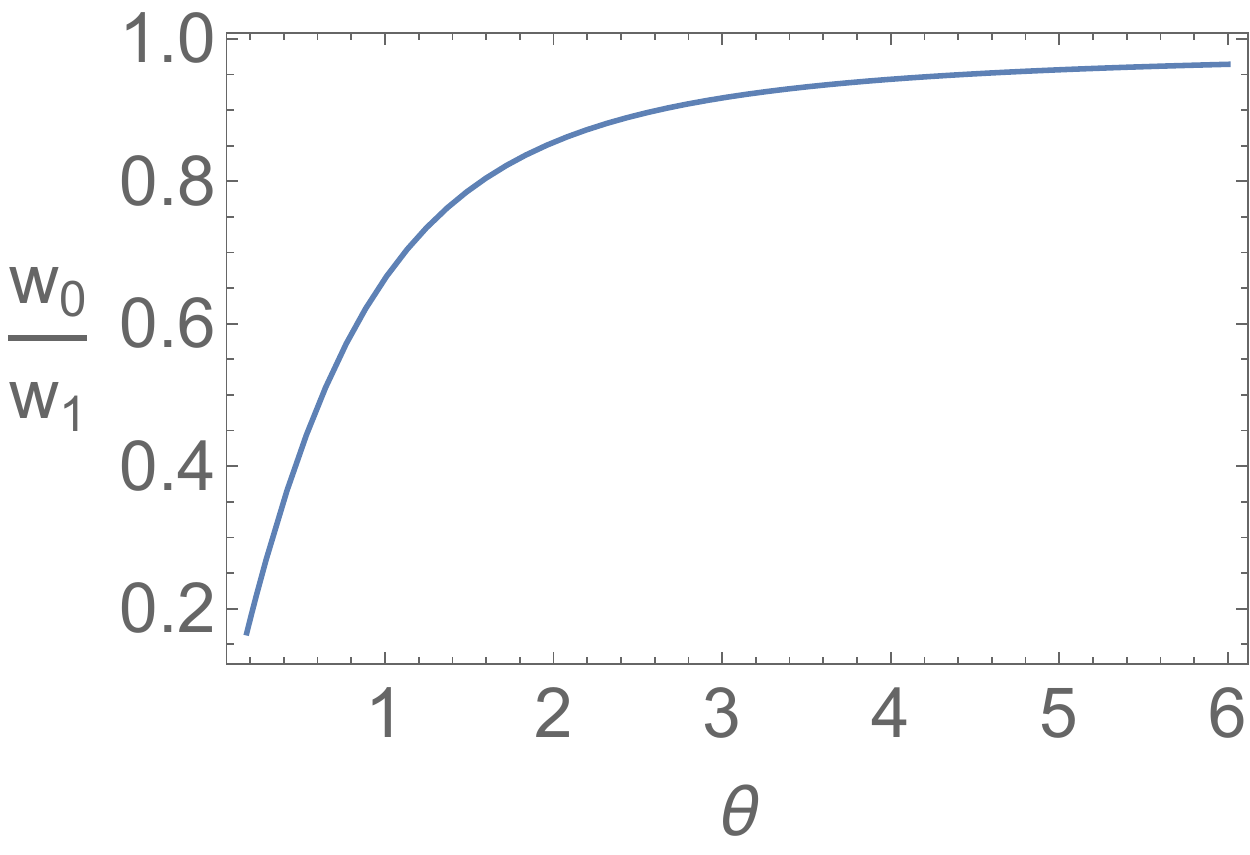}
	\end{center}
	\caption{(Color online) Plot of the ratio $w_0/w_1$ as function of angle (in degree) for the range of its validity.}.
	\label{fig:w0overw1}
\end{figure}

This also means that the lowest Landau level at small twist angles is split slightly less dominantly than in the unrelaxed $w_1=w_0$ case. Therefore, for a relaxed lattice one might need a slightly higher (than the $w_0=w_1$ case) experimental energy resolution to resolve the individual energy levels arising from the lowest Landau level. However, to be fair, this effect is relatively minuscule. For instance at $\theta=1^\circ$, $w_1=110$ meV and $p/q=0.05$ the width of the split lowest Landau level is reduced by only 10\%. Therefore the additional energy resolution that is needed to resolve the individual levels coming from the lowest Landau level can be estimated at 10\% higher than for the case $w_0=w_1$. Nevertheless, it should be stressed that this effect might potentially be more dominant in other moir\'{e} materials.

\section{Circularly polarized light}
\label{Circularly polarized light}
In this section we will turn our attention to the effect that circularly polarized light has on the Hofstadter butterfly in TBLG.
\subsection{Effective Hamiltonian}
If perpendicularly incident circularly polarized light (we consider the right-handed or clock-wise case) is applied to the graphene layers, at frequency $\Omega$ and driving strength $A$, it enters purely via minimal substitution as \cite{Vogl1,Dehghani,Ibsal}
\begin{equation}
 \tilde{k}_{x}(t)=k_{x}-A \cos (\Omega t), \quad \tilde{k}_{y}(t)=
k_{y}-A \sin (\Omega t).
\end{equation}
Thus, we have a time-periodic Hamiltonian satisfying  $H(\mathbf{x}, \mathbf{k},t+2\pi/\Omega)=H(\mathbf{x}, \mathbf{k},t)$.
It is a well-known fact that some of the physical features of a periodically driven Hamiltonians can be approximately captured by an effective time independent Hamiltonian \cite{Vogl1}. In our case such a description will be advantageous for numerical reasons. Therefore, let us briefly recall how to arrive at an effective time independent description and what new physical effects are introduced.

A non-perturbative scheme to find effective time-independent Hamiltonians for a periodically driven system is to transform its Hamiltonian to a rotating frame (RF) $H_R =U^{\dagger}(t)\left(H-i \partial_{t}\right) U(t)$ via a unitary transformation \cite{Vogl1}. If a convenient frame is chosen a subsequent time average generates a Hamiltonian that is more accurate than Hamiltonians found via the usual high frequency approximations such a van Vleck or Floquet-Magnus \cite{Vogl4}. With such a properly chosen unitary transformation it was found \cite{Vogl1,Ibsal} that a highly accurate effective Hamiltonian for TBLG subject to a  circularly polarized light is given by
\begin{equation}
	H(\mathbf{x}, t)=\begin{pmatrix}
		h(\theta_1, \mathbf{k}) & \tilde{T}(\mathbf{x}) \\
		\tilde{T}^{\dagger}(\mathbf{x}) & 	h(\theta_2,  \mathbf{k})
	\end{pmatrix}.
	\label{eq:TBG_RotFrame}
\end{equation}
It is important to stress that the derivation from \cite{Vogl1,Ibsal} generalizes directly to a case with a magnetic field because the unitary transformations that were used are momentum-independent.
In the expression \eqref{eq:TBG_RotFrame} above we observe that the intralayer Hamiltonians have been modified by light as follows  
\begin{equation}
	h(\theta, \mathbf{k})=v_{\mathrm{RF}} R(\theta) \mathbf{k}\cdot \boldsymbol{\sigma}-\Delta_{\mathrm{RF}} \sigma_{3},
\end{equation}
where again we recall that $R(\theta)$ is a rotation matrix in the layer plane.  
The Fermi velocity is also modified and becomes
\begin{equation}
	v_{\mathrm{RF}}=v_F J_{0}\left(-\frac{6 \gamma}{\Omega} J_{1}\left(\frac{2 A a_{0}}{3}\right)\right) J_{0}\left(\frac{2 A a_{0}}{3}\right)
\end{equation}
where  $J_0$ is the zeroth Bessel function of the first kind. Hereby $Aa_0$ provides a unitless scale of driving strengths. 
Light also causes the system to acquire a band gap, which is given by
\begin{equation}
\Delta_{\mathrm{RF}}=-\frac{3 \gamma}{\sqrt{2}} J_{1}\left(\frac{2 A a_{0}}{3}\right) J_{1}\left(-\frac{6 \sqrt{2} \gamma}{\Omega} J_{1}\left(\frac{2 A a_{0}}{3}\right)\right).
\end{equation}
Interlayer tunneling matrices are also modified. If we express $T_{j}=\sum_i T_{j,i}\sigma_i$, where $T_{j,i}$ are expansion coefficients. Then modified interlayer hopping matrices $\tilde{T}_{j}$ are given by
\begin{equation}
    \tilde{T}_{j}=\sum_i T_{j,i}\tilde \sigma_i
\end{equation}
where $\tilde{\sigma}_{1,2}=J_{0}(\nu) \sigma_{1,2}$ and
\begin{equation}
	\begin{aligned}
		&\tilde{\sigma}_{0}=\sigma_{0}+\left(J_{0}(\sqrt{2} \nu)-1\right)\left[\sigma_{0} \sin ^{2}\left(\frac{\theta}{2}\right)+\frac{i}{2} \sigma_{3} \sin \left(\theta\right)\right]\\
		&\tilde{\sigma}_{3}=\sigma_{3}+\left(J_{0}(\sqrt{2} \nu)-1\right)\left[\sigma_{3} \cos ^{2}\left(\frac{\theta}{2}\right)-\frac{i}{2} \sigma_{3} \sin \left(\theta\right)\right]
	\end{aligned}
\end{equation}
with $ \nu=(-6 \gamma / \Omega) J_{1}\left(2 A a_{0} / 3\right)$. 

We may now include the magnetic field in the Landau gauge via minimal substitution and use a convenient basis like in Eq. \eqref{eq:magn_basis} to express the Hamiltonian in a numerically advantageous form. It is then found that the expressions are almost the same as Eqs. \eqref{eq:htheta} and \ref{eq:Thopmagbas} just with $\omega_c\to \omega_{RF}=\omega_c v_{RF}/v_F$ and $T_i\to \tilde T_i$, of course with an additional gap term 
\begin{equation}
		\eqfitpage{H_\Delta=\Delta_{RF}\sum_{L, n, y}\paren[\bigg]{\ket{L, n, A, y}\bra{L, n, A, y}-\ket{L, n, B, y}\bra{L, n, B, y}}},
		\label{eq:hdelta}
\end{equation}
which we use in our numerical analysis.
\subsection{Numerical Results}
\label{NUMERICAL RESULTS AND DISCUSSION}

\begin{figure*}
	\centering
	\includegraphics[width=0.39\linewidth]{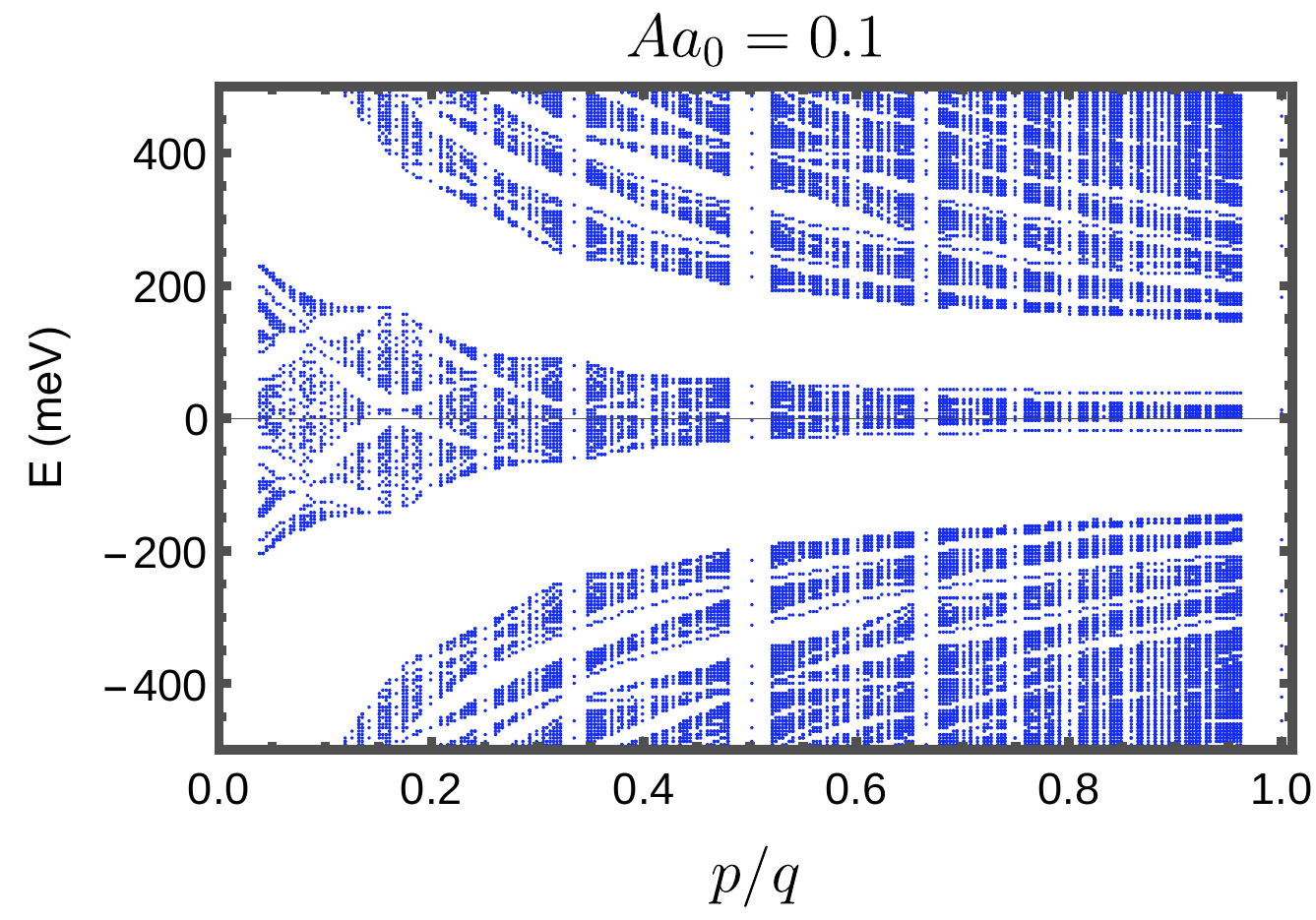}
	 \includegraphics[width=0.39\linewidth]{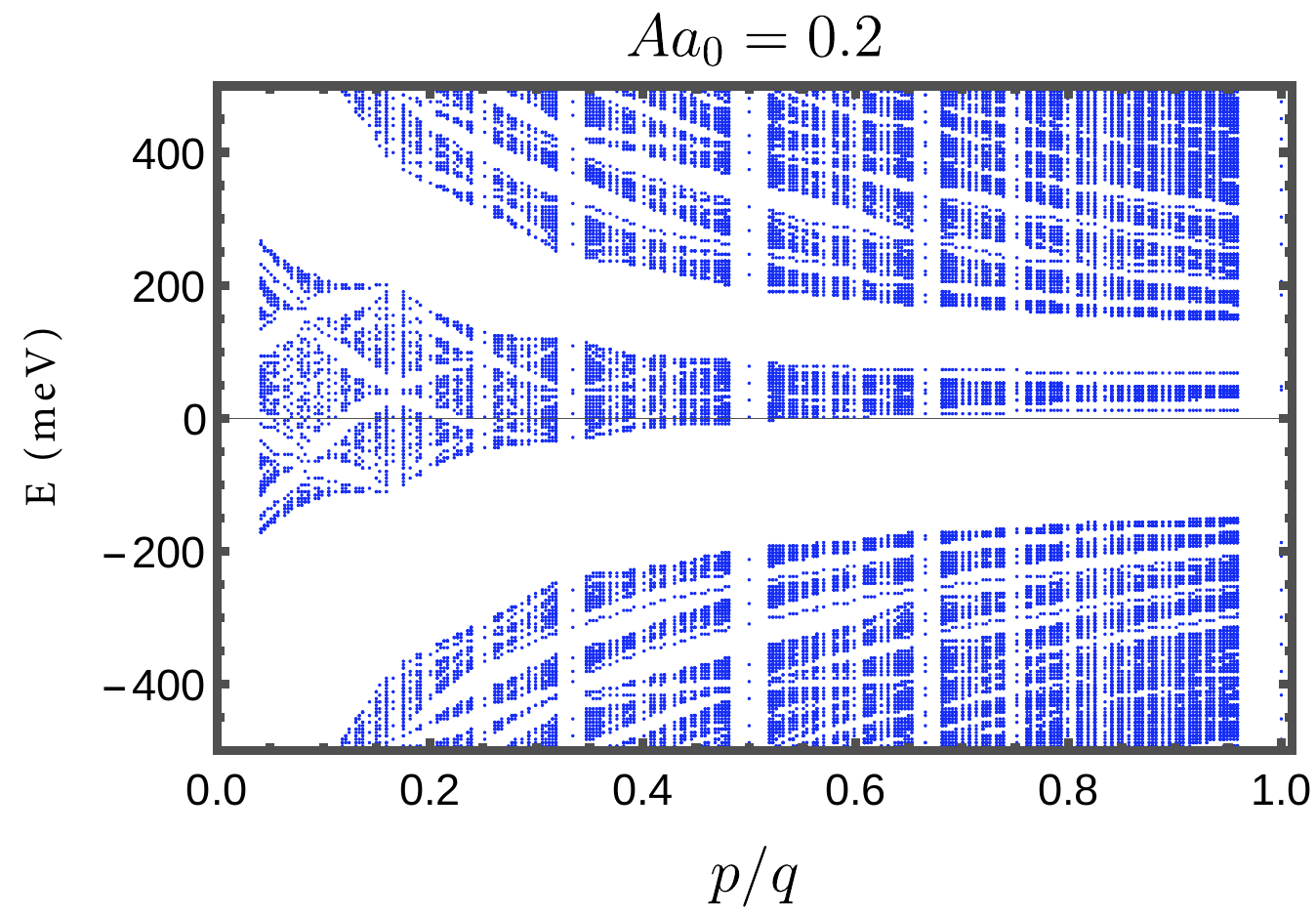}\\
	\includegraphics[width=0.39\linewidth]{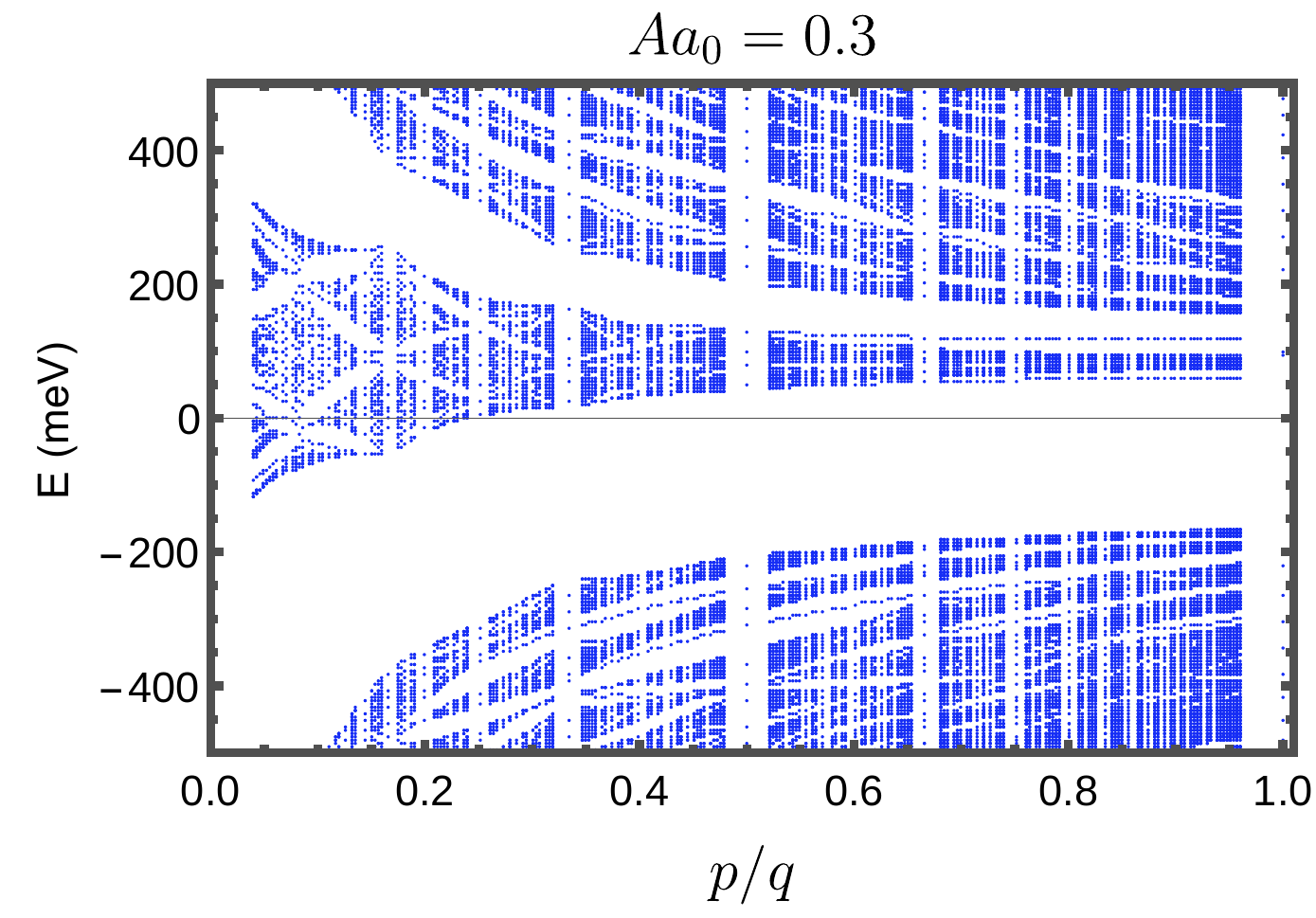}
	 \includegraphics[width=0.39\linewidth]{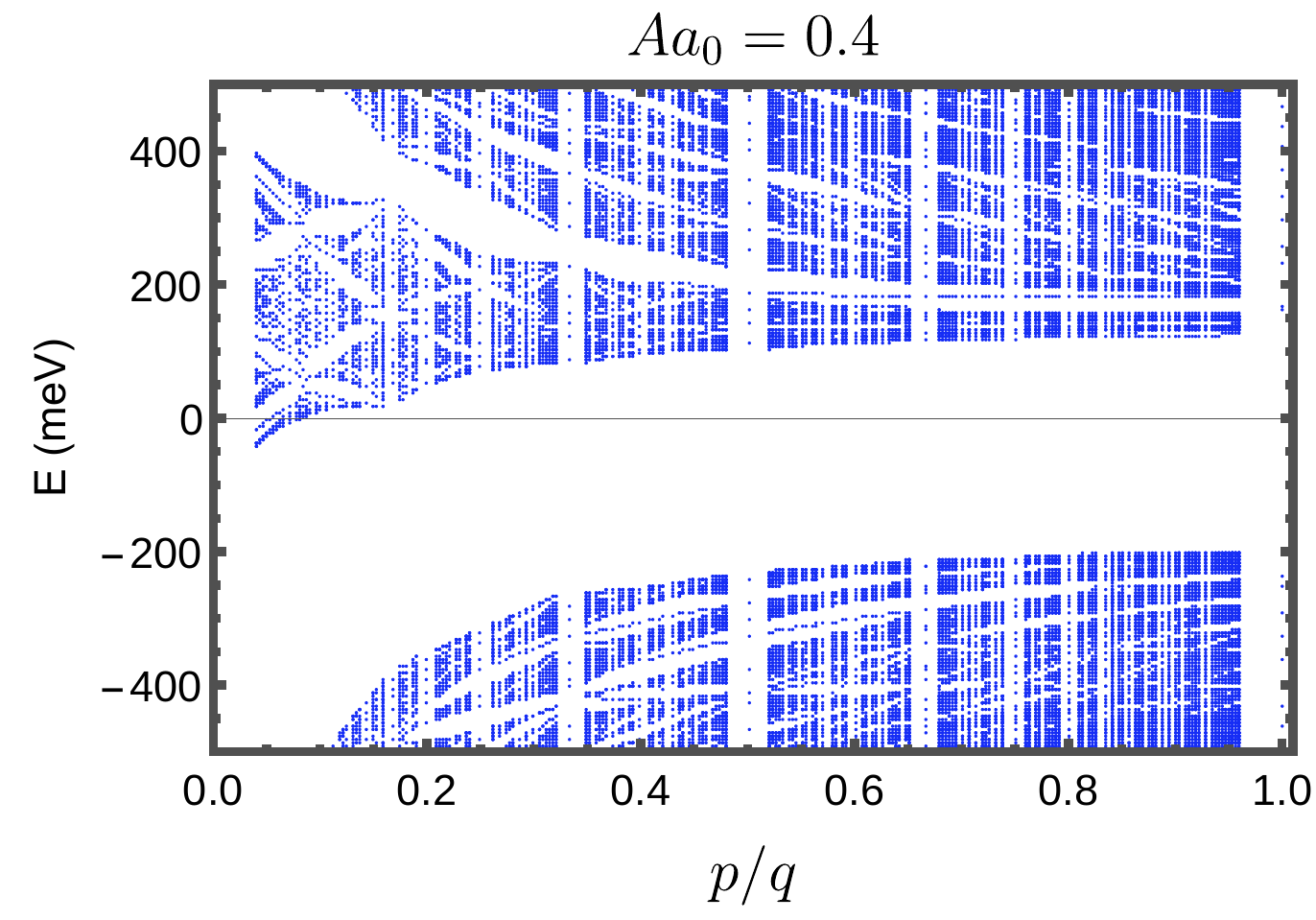}
	\caption{(Color online) The Floquet Hofstadter butterfly spectrum as function of $p/q=\phi_0/\Phi$, subject to right-handed circularly polarized light with driving frequency fixed at $\Omega=2\gamma$. The representative driving strengths are chosen as $Aa_0 = 0.1$, $Aa_0 = 0.2$,  $Aa_0 = 0.3$,  $Aa_0 = 0.4$. The parameters used are, $\gamma = 2364$ meV, $w_0=0.9w_1$, $w_1=110 $ meV and $\theta=2^\circ$. The Landau level cut-off that was used for the plots is the same as the one suggested in \cite{ MacDonlad} $n_{\mathrm{max}} \approx 2\left[\max \left(a_{0}\gamma_{\text{RF}}, w_1\right) / \omega_{c}\right]^{2}$}.
	\label{fig:circ_pol_light_butterfly2gamma}
\end{figure*}

We will now numerically investigate the effect that circularly polarized light has on the Hofstadter butterfly. For this we have plotted the Hofstadter butterfly at a fixed driving frequency of $\Omega=2\gamma$ (chosen to be in the high frequency regime) and various driving strengths, which can be seen in Fig. \ref{fig:circ_pol_light_butterfly2gamma} below 
We observe that as the driving strength is increased, the asymmetry of the energy levels with respect to $E=0$ becomes increasingly apparent. 
Now it is interesting to also consider the cases of either $w_0=0$ or $w_1=0$ that have a spectrum which in the equilibirum case was symmetric with respect to $E=0$.
In both cases we find that energy levels corresponding to the levels that appear from the split 0-th Landau levels of two decoupled graphene layers (we  will call it center branch of the butterfly) move upwards to higher energies as we increase the driving strength $Aa_0$. This introduces an apparent asymmetry of the spectrum with respect to $E=0$. The source of this asymmetric behaviour is obvious because the chiral symmetry we discussed earlier is broken by the introduction of term $\Delta_{\mathrm{RF}}$.

\begin{figure}
	\begin{center}
	\includegraphics[width=0.8\linewidth]{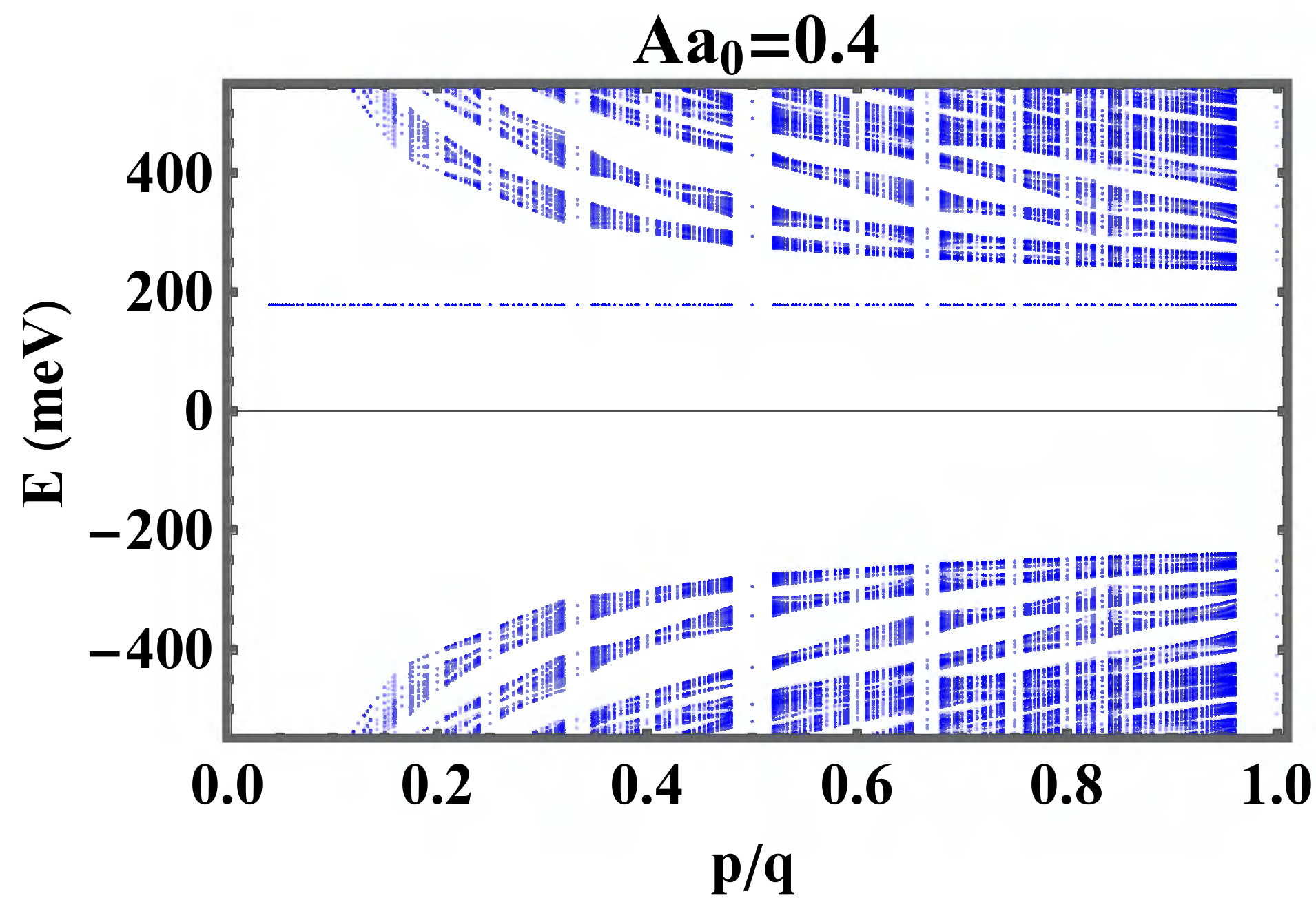}
	 \includegraphics[width=0.8\linewidth]{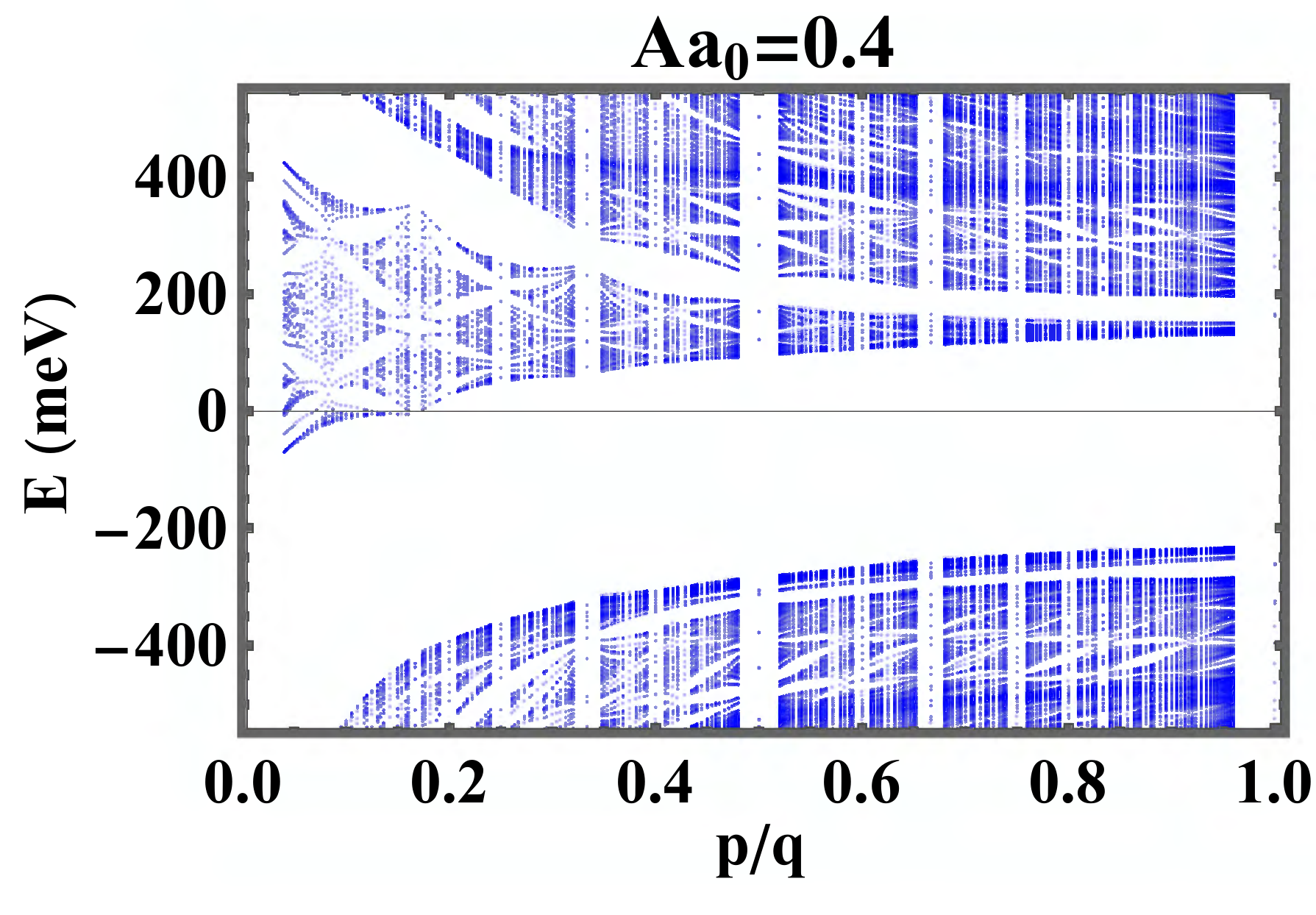}
	\end{center}
	\caption{(Color online) The Floquet Hofstadter butterfly spectrum as function of $p/q=\phi_0/\Phi$, subject to right-handed circularly polarized light with driving frequency fixed at $\Omega=2\gamma$, where  $\gamma = 2364$ meV. A representative driving strength is chosen as  $Aa_0 = 0.4$ and the angle was set as $\theta=2^\circ$. The top figure corresponds to the case $w_0=0$, $w_1=110 $ meV and bottom figure $w_0=110$, $w_1=0 $ meV. The Landau level cut-off that was used for the plots is the same as the one suggested in \cite{ MacDonlad} $n_{\mathrm{max}} \approx 2\left[\max \left(a_{0}\gamma_{\text{RF}}, w_1\right) / \omega_{c}\right]^{2}$}.
	\label{fig:wssettozerocircpol}
\end{figure}

However, interestingly in the case of $w_0=0$ and $w_1\neq 0$ it is found that this asymmetry appears only for the center branch of the butterfly - all other levels, up to numerical accuracy, remain symmetric with respect to $E=0$. In contrast for the case of $w_1=0$ and $w_0\neq 0$ we find that the complete spectrum becomes asymmetric with respect to $E=0$ so there seem to be no remnants of a chiral symmetry. Both cases can be seen in Fig. \ref{fig:wssettozerocircpol}.

Finally, it is important to mention that there is an easy way to understand why the center branch of the Hofstadter butterfly moves upward - rather than downward. Particularly, this can be understood from the Landau levels of single graphene layer, which are just split into the branches of our butterfly. Let us first consider two isolated graphene layers. Here, for each layer the wavefunction near the K point has the form $\psi_n=(\pm\ket{n-1},\ket{n})$. In the case of $n=0$ one should note that there is only a lower component. For each layer, the gap term $\Delta_{\mathrm{RF}}$ enters as $-\Delta_{\mathrm{RF}}\sigma_z$. To leading order perturbation theory we can then directly see that this shifts the $n=0$ Landau level upwards. It may seem surprising that symmetry is broken in this way. We should also point out that while near the $K^\prime$ point the opposite shift $\Delta_{\mathrm{RF}}\sigma_z$ happens, the center band is actually shifted in the same direction as in the case of the $K$ point. This is because for the $K^\prime$ point $\psi_n\to(\ket{n},\pm\ket{n-1})$, i.e. the first and second components are flipped, which compensates for the change in the sign of $\Delta_{\mathrm{RF}}$. A plot showing this can be found in appendix \ref{app:kprime}. For the remainder of this work we focus on the physics near the K points of the graphene layers because for our purposes the physics near the $K^\prime$ point does not differ substantially from the physics near the $K$ point.

\begin{figure}
	\begin{center}
	\includegraphics[width=0.8\linewidth]{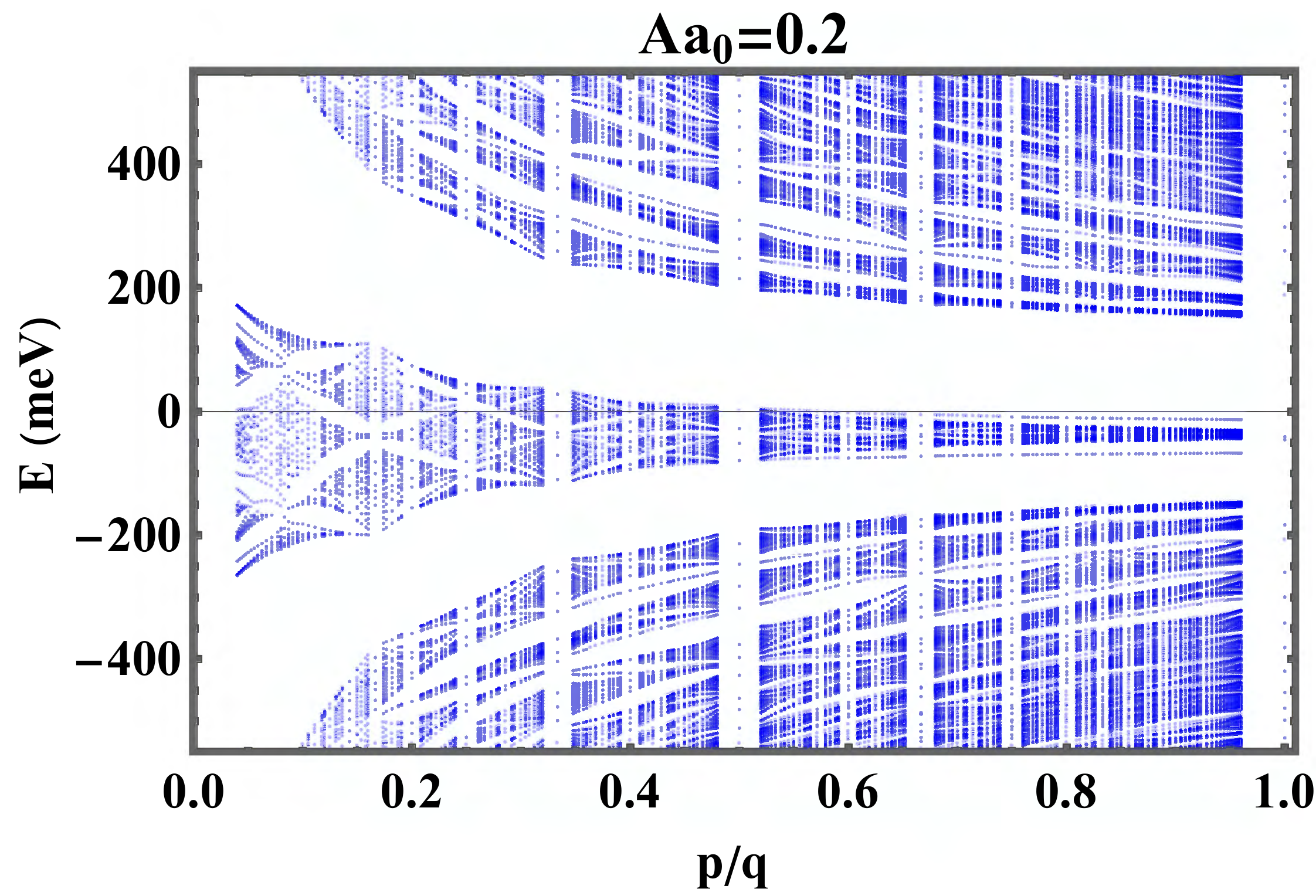}
	\end{center}
	\caption{(Color online) The Floquet Hofstadter butterfly spectrum as function of $p/q=\phi_0/\Phi$, subject to left-handed circularly polarized light with driving frequency fixed at $\Omega=2\gamma$. A representative driving strength was as  $Aa_0 = 0.2$. The parameters used are, $\gamma = 2364$ meV, $w_0=0.9w_1$, $w_1=110 $ meV and $\theta=2^\circ$. The Landau level cut-off that was used for the plots is the same as the one suggested in \cite{ MacDonlad} $n_{\mathrm{max}} \approx 2\left[\max \left(a_{0}\gamma_{\text{RF}}, w_1\right) / \omega_{c}\right]^{2}$}.
	\label{fig:circ_pol_lefthand}
\end{figure}

Lastly, we may consider the question on how to obtain a center band that is shifted downward, rather than upward. This can be answered quite easily via left handed circularly polarized light, which can be seen in Fig. \ref{fig:circ_pol_lefthand}.

We see that the center Landau level is shifted downward just as we wanted. This can be understood quite easily. Specifically, to leading order in Floquet theory it is found easily that the main effect of the change from right-handed to left-handed circularly polarized light is the replacement $\Delta_{RF}\to -\Delta_{RF}$. That is, the lowest Landau level is shifted downward rather than up like in the case of right-handed circularly polarized light.

\section{Waveguide light}
\label{Waveguide light}
After our discussion of circularly polarized light, in this next section we will focus on the effects due to a second type of light, a linearly polarised light emanating from a waveguide.
\subsection{Theoretical Approach}
The second type of light that will be considered  is longitudinal light coming from a waveguide. Here, the boundary conditions of a waveguide allow for light with longitudinal components $\vect A=\mathrm{Re}(e^{ik_z z-i\Omega t})\hat z$ to exist, which is not possible in a vacuum - this expression is valid in a carefully chosen spatial region (more details can be found in the appendix \cite{Vogl2} or most standard references on electromagnetism such as \cite{JDJACKSON}).
The effect of this light can be included within the tight binding model via a Peirls substitution $t_{ij}\to t_{ij}\exp(-\int_{\vect r_i}^{\vect r_j} d\vect l \vect A)$. In the continuum Hamiltonian interlayer hopping terms correspond to $w_i$, which is why the $w_i\to w_i \exp(-\int_{\vect r_i}^{\vect r_j} d\vect l \vect A)$. To leading order in the high frequency regime of our continuum model (that is the average of the Hamiltonian over one period) this effect can be captured if we replace interlayer couplings as given below
\begin{equation}
w_{0} \rightarrow w_{0} J_{0}\left(A a_{\text{AA}}\right), \quad w_{1} \rightarrow w_{1} J_{0}\left(A a_{\text{AB}}\right).
\label{eq:wireplacements}
\end{equation}
More details on the derivation can be found in \cite{Vogl2}. We note that in \cite{Vogl2} it is also found that this lowest order approximation captures most of the features of a more detailed treatment that uses an extended space approach. We therefore restrict our attention to this case.
Hereby, $a_{\text{AA}} = 0.36$ nm and $a_{\text{AB}} = 0.34$ nm are interlayer distances in $\text{AA}$ and $\text{AB}$ regions of TBG. We should note that with the interlayer spacing we made the assumption that AA and AB stacked regions are well pronounced enough to have the corresponding lattice spacing. This is a simplification of the model that becomes well justified in the case of very small twist angles where AB stacked regions become increasingly pronounced due to lattice relaxation effects. 
\subsection{Numerical Results}

We will now discuss the effect of the waveguide light on the Hofstadter butterfly via numeric results. Since the effect that waveguide light has on the Hofstadter butterfly does not differ considerably between the $K$ and $K^\prime$ points, we chose to only consider the case of the $K$ point. 

Here, we consider a range of  different values for our unit-less driving strength $Aa_{\text{AA}}$. Note that
\begin{equation}
 Aa_{\text{AB}}=\frac{a_{\text{AB}}}{a_{\text{AA}}}Aa_{\text{AA}}.
\end{equation}

We consider different values for the driving strength in the range $Aa_{\text{AA}}=0.2$ to $4$ in the Fig.  \ref{fig:hofstadter_waveguide_light} below.
\begin{figure*}[ht]
	\centering
	\includegraphics[width=0.39\linewidth]{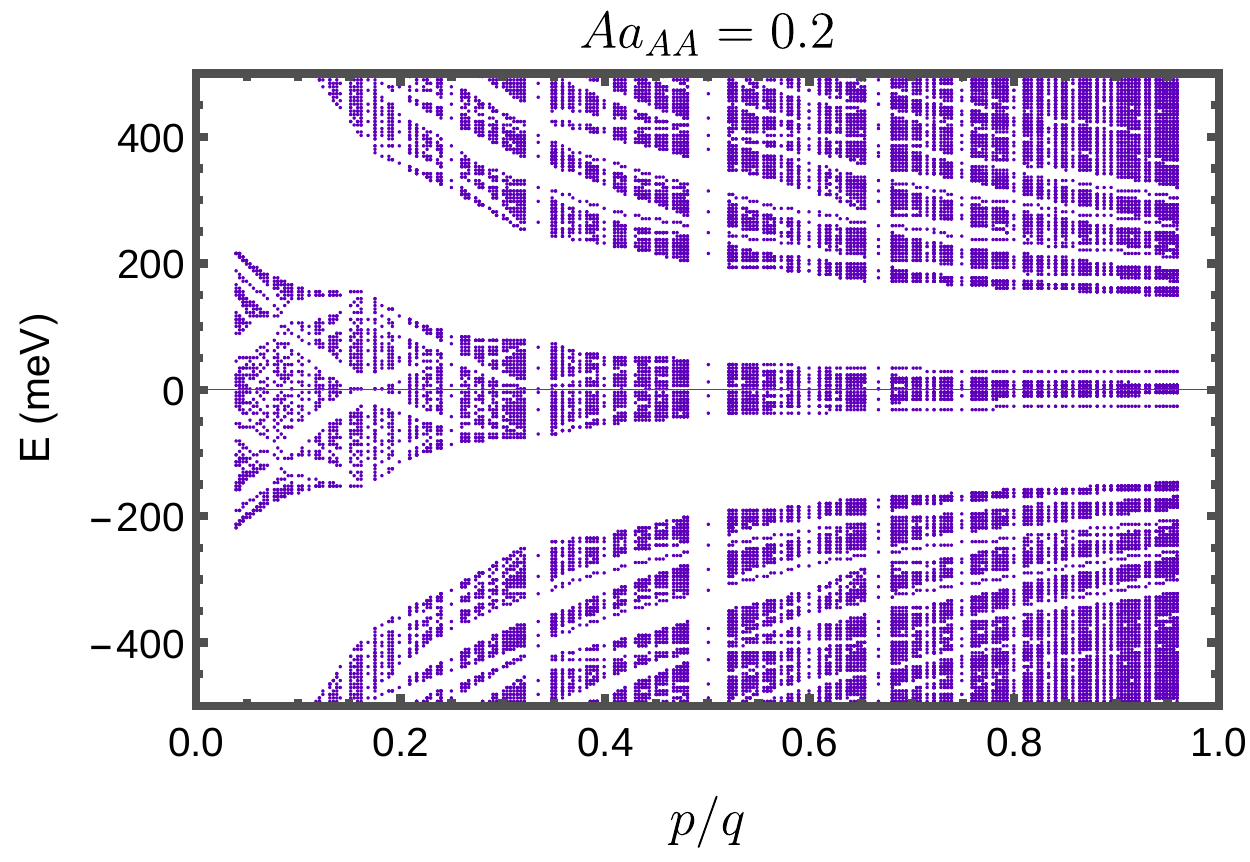}
	\includegraphics[width=0.39\linewidth]{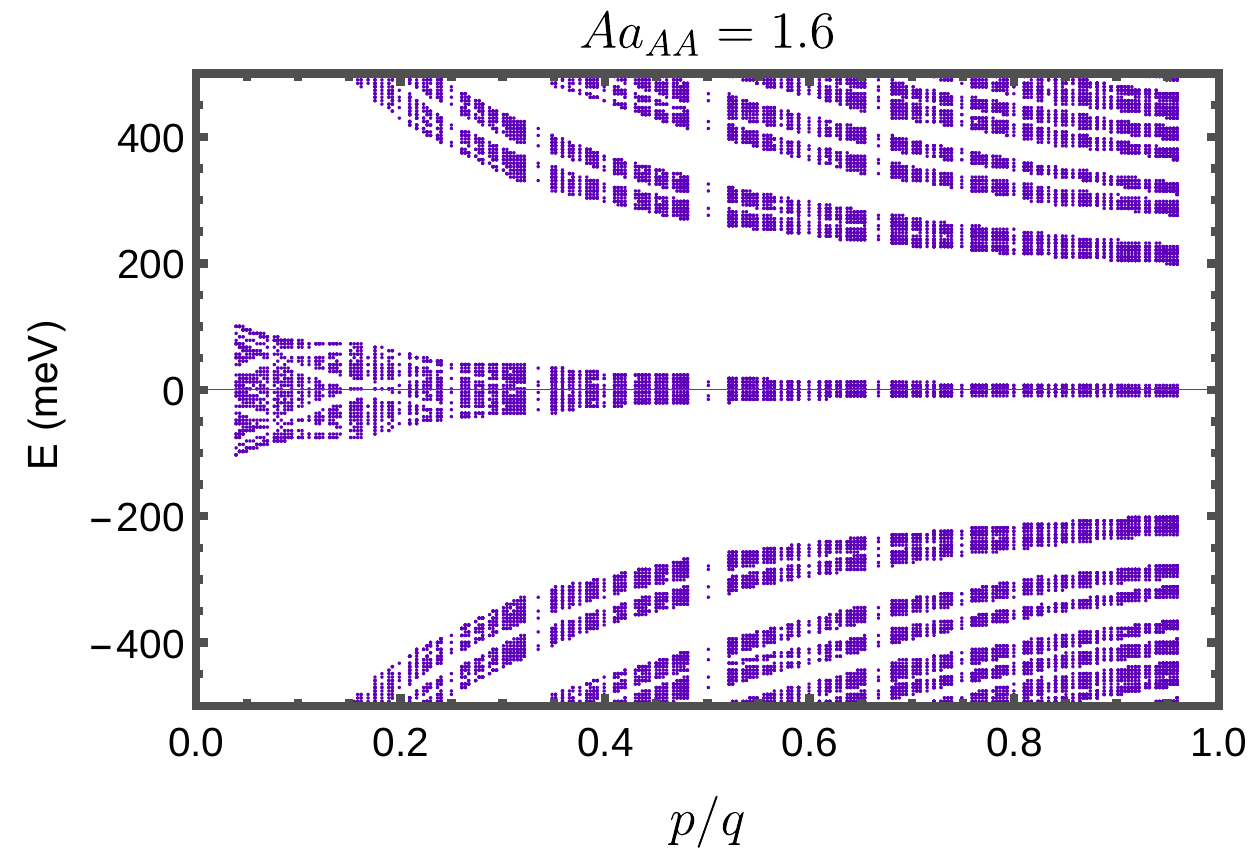}
	\includegraphics[width=0.39\linewidth]{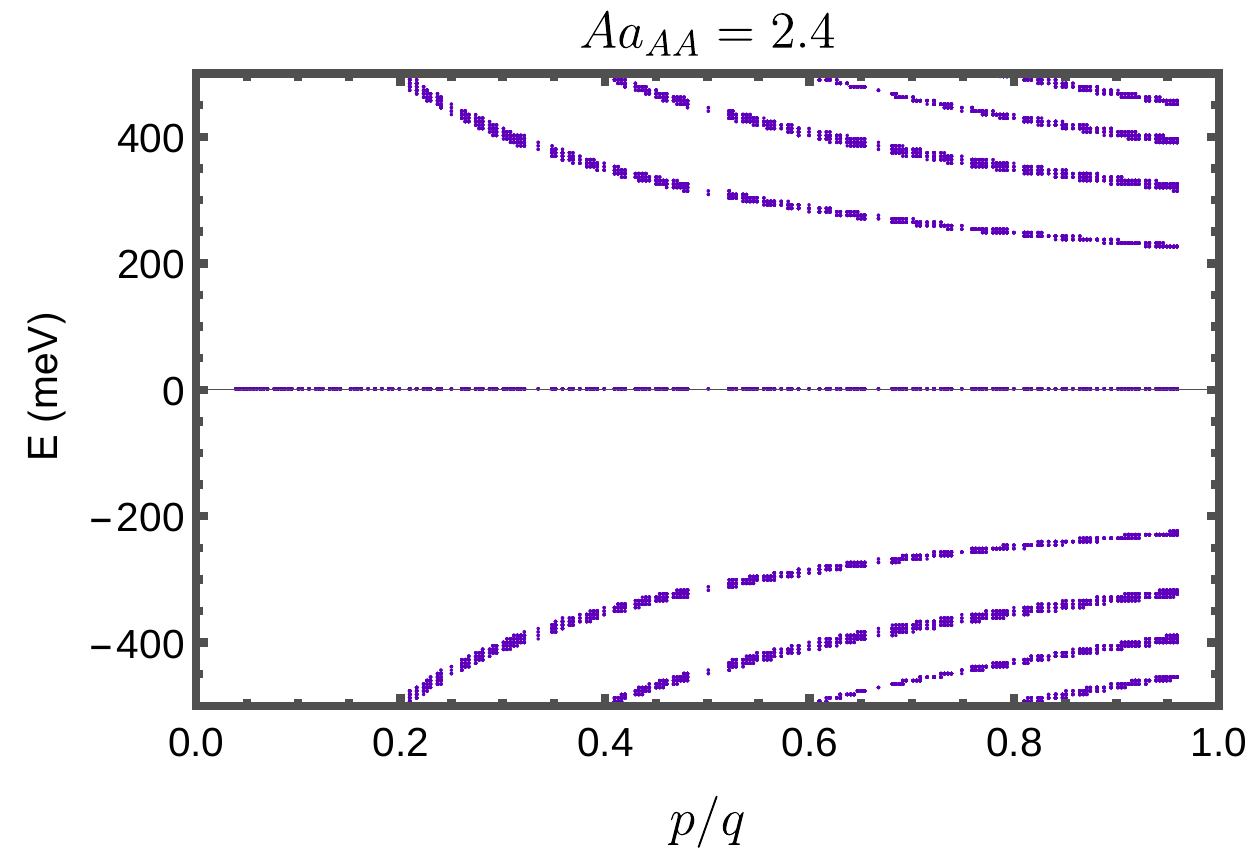}
	\includegraphics[width=0.39\linewidth]{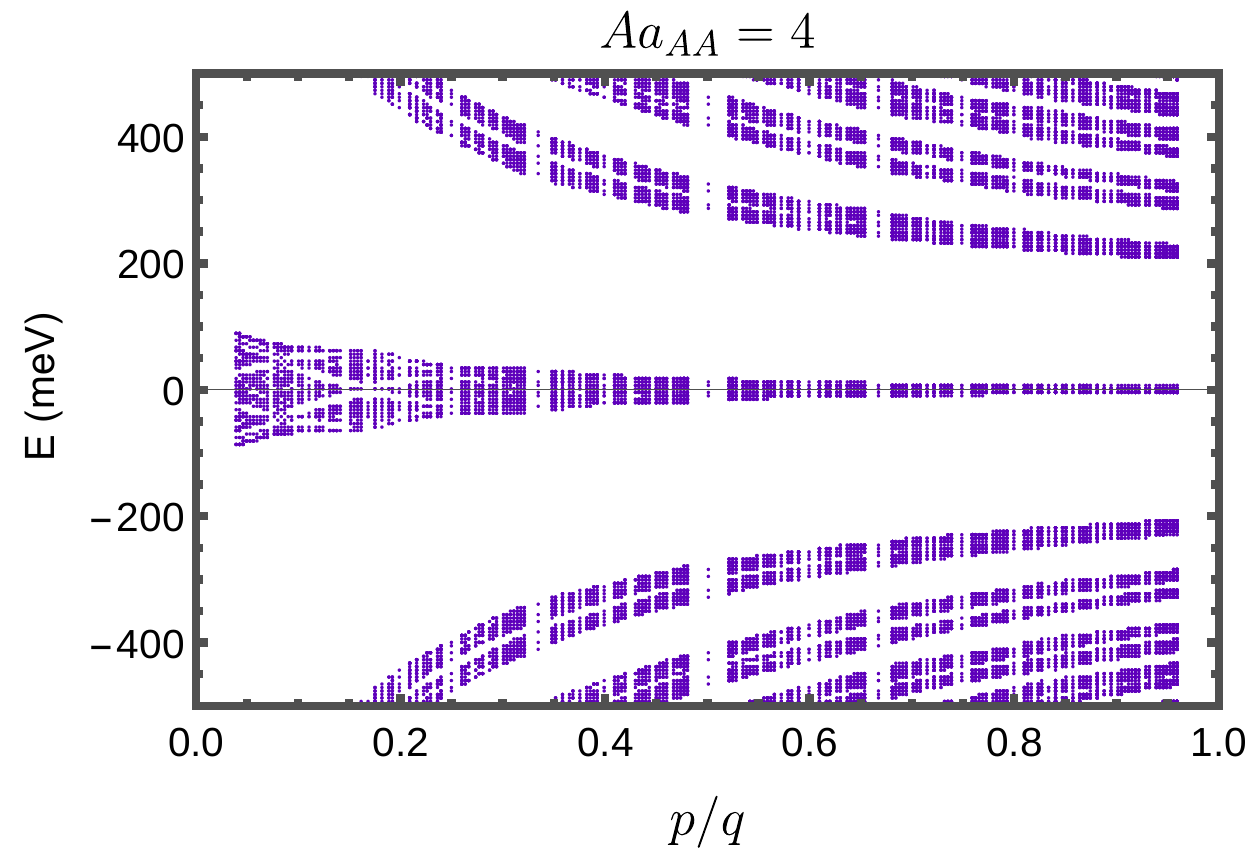}
	\caption{(Color online) The Floquet Hofstadter butterfly spectrum as function of $p/q=\phi_0/\Phi$, subject to waveguide light. The representative driving strengths are chosen from $Aa_{\text{AA}} = 0.2$ to $Aa_{\text{AA}} =4$. The parameters used are, $\gamma = 2364$ meV, and $\theta=2^\circ$.}\label{fig:hofstadter_waveguide_light}
\end{figure*}
We find that the level splitting of the individual Landau levels first decreases as we increase $Aa_{\text{AA}}$ and then eventually increases again. This can be explained very easily using the fact that $w_0$ and $w_1$ set the scale for the size of level splitting, which in our case are modulated by Bessel functions $J_0$. 
This observation of course allows us to go one step further to find that the two chiral models either $w_1=0$ or $w_0=0$ can be realized via this form of light. Particularly, $w_0$ is effectively set to zero if $Aa_{\text{AA}}=j_{0,n}$ is the $n$-th zero of the Bessel function $J_0(x)$. Similarly $w_1$ is effectively set to zero if $Aa_{\text{AA}}=(a_{\text{AA}}/a_{\text{AB}})j_{0,n}$. At these points the mirror symmetry of the spectrum with respect to $E=0$ is restored. 

We should stress regarding the appropriateness of the approximation in Eq. \eqref{eq:wireplacements}. When the band structure resulting from this approximation was compared to band structures resulting from a full extended space treatment of the time dependent problem such as in \cite{Vogl2} it was found that even at driving strengths as high as $\eta=Aa_{\text{AA}}=4$ the approximation yielded results that were almost indistinguishable from the exact extended space treatment. 

Last, but not the least, regarding the potential experimental realizability of such large driving strengths. We can recall that (reintroducing units) $\eta=\frac{eEa_{AA}}{\hbar \Omega}$ (see \cite{Vogl2}). Here, for driving frequencies in the high frequency regime $\hbar\Omega>2.7$ eV and electric field strengths up to $E<15$ meV/cm, $\eta<0.2$ is rather limited. Here, the high frequency assumption $\hbar\Omega>2.7$ eV is our main limiting factor. However, Landau levels are flat and therefore one can find regimes of $p/q$ and $\theta$ where $\hbar\Omega<100$ meV is a high frequency regime  for the center part of the Hofstadter butterfly (see e.g. Fig. \ref{fig:TBG_combined_figc} to identify such a regime visually - being careful that for the high frequency regime the driving strength also needs to be smaller than the gap size). Therefore our high frequency approximation can be justified with certain restrictions while values of $\eta>4$ are achievable experimentally. Therefore, both chiral limits of TBLG should be achievable experimentally by employing waveguide light.

Lastly, we consider the case where one starts from one of the chiral limits $w_0=0$ or $w_1=0$. Here, we find that this form of light does not lead to chiral symmetry breaking, in contrast to the circularly polarized light case. 
\FloatBarrier
\section{Summary and conclusion}
\label{Summary and conclusion}
We have studied the Floquet Hofstadter butterfly spectrum in TBLG subject to a uniform perpendicular magnetic field both in the equilibrium case and in the presence of different forms of light. We have focused on the cases of circularly polarized light and waveguide linearly polarized light. For the equilibrium case we have identified two separate chiral limits and found that one of them - $w_0\neq 0$ and $w_1=0$ - includes most of the butterfly physics. 

In the case of circularly polarized light, we have found that the main effect is the creation of a gap term that breaks chiral symmetry. This causes the Hofstadter butterfly to deform such that levels are visibly and based on numerical grounds less mirror symmetric with respect to $E=0$. In the case of waveguide light we have found that it does not break chiral symmetries if they are present from the start. Interestingly, field driving strength can be selected to achieve both of TBLG's chiral limits. That is, we have found a way to design an experimentally accessible regime that allows the realization of the two chiral limits of TBLG.

\section*{Acknowledgments} H.B. and M.V. gratefully acknowledge the support provided by the Deanship of Research Oversight and Coordination (DROC) at King Fahd University of Petroleum \& Minerals (KFUPM) for funding this work through start up project No.SR211001.

	\appendix
	\newpage
	\section{Discussion of the $K^\prime$ point}
	\label{app:kprime}
	For the reader's convenience we have also included in this section a brief discussion of the $K^\prime$ point. First we note that the Hamiltonian in the Landau level basis can be obtained if we make the following replacements, which are equivalent to the replacements that were mentioned in the main text.
	
	\begin{equation}
	    \begin{aligned}
	    &\eqfitpage{h(\theta/2)\to \omega_{c} \sum\limits_{L, n, j}\left(e^{-i\theta/2} \sqrt{n+1}\ket{L, n+1, B, j}\bra{L,n, A, j}+\mathrm{H.c.}\right)}\\
	    &\vect q_i\to-\vect q_i;\quad \zeta\to\zeta^*;\quad \Delta_{RF}\to-\Delta_{RF};\quad \theta\to-\theta
	    \end{aligned}
	\end{equation}
	
	We stress that one has to be careful about the order of these replacements and apply them in the listed order. Another subtle point one has to be mindful about is that one has to ensure the correct type of sublattice symmetry breaking through the choice of basis set - the lowest Landau level state at each K or $K^\prime$ point does not have this symmetry. Therefore, to ensure a numerical error that happens only at large energies one had to introduce basis sets for A and B sublattices that are altered from the main text but serve the same purpose, which is to remove spurious low energy states.  Recall that $(\pm \ket{n-1},\ket{n})\to(\pm \ket{n},\ket{n-1})$. Particularly our choice of basis has to be according to
	\begin{equation}
	    \{L\in\{t,b\},\alpha\in\{A,B\},n\in\{0,\dots,n_{\mathrm{max}}-\delta_{\alpha,A}\}\},
	\end{equation}
	where $L$ is the layer degree of freedom, $\alpha$ the sublattice and $n$ the Landau level index. The term $\delta_{\alpha,A}$ ensures that the Landau level index is truncated earlier for sublattice A.

\begin{figure}
	\begin{center}
	\includegraphics[width=0.8\linewidth]{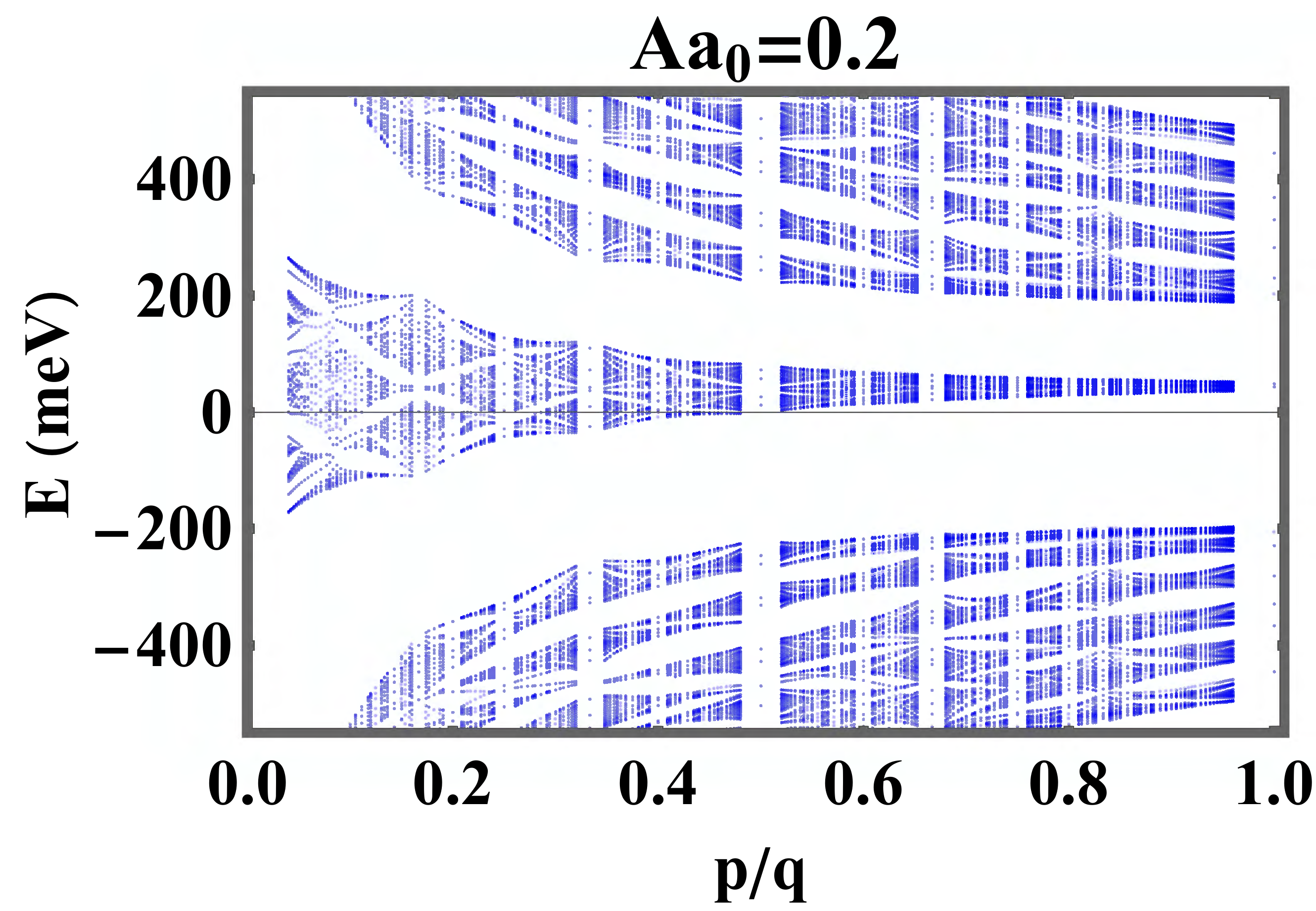}
	\end{center}
	\caption{(Color online) The Floquet Hofstadter butterfly spectrum as function of $p/q=\phi_0/\Phi$, subject to right-handed circularly polarized light with driving frequency fixed at $\Omega=2\gamma$. A representative driving strength was chosen as $Aa_0 = 0.2$. The parameters used are, $\gamma = 2364$ meV, $w_0=0.9w_1$, $w_1=110 $ meV and $\theta=2^\circ$. The Landau level cut-off that was used for the plots is the same as the one suggested in \cite{ MacDonlad} $n_{\mathrm{max}} \approx 2\left[\max \left(a_{0}\gamma_{\text{RF}}, w_1\right) / \omega_{c}\right]^{2}$. The plot is valid near the $K^\prime$ point. The top right of Fig. \ref{fig:circ_pol_light_butterfly2gamma} is a comparable figure for the K point}.
	\label{fig:circ_pol_light_butterfly2gamma2}
\end{figure}

	With these changes to the Hamiltonian we were able to generate plots that are valid near the $K^\prime$ point. The presumably most interesting situation occurs in the case of circularly polarized light with a plot in Fig. \ref{fig:circ_pol_light_butterfly2gamma2} given below

We see that even for this case there is no major physics that differs fundamentally from the physics near the $K$ point, this validates our decision to consider only the $K$ point in our main text.
	

\begin{thebibliography}{99}
\bibitem{Tarnopolsky} G. Tarnopolsky, A. J. Kruchkov, and A. Vishwanath, Origin of Magic Angles in Twisted Bilayer Graphene, \href{https://doi.org/10.1103/PhysRevLett.122.106405} {Phys. Rev. Lett. \textbf{122}, 106405 (2019)}.
\bibitem{Novoselov} K. S. Novoselov, A. K. Geim, S. V. Morozov, D. Jiang, Y. Zhang, S. V. Dubonos, I. V. Grigorieva and A. A. Firsov, Electric field effect in atomically thin carbon films, \href{https://doi.org/ 10.1126/science.1102896}{Science \textbf{306}, 666 (2004)}.	
\bibitem{CastroNeto} A. H. Castro Neto, F. Guinea, N. M. R. Peres, K. S. Novoselov, and A. K. Geim, The electronic properties of graphene, \href{https://doi.org/10.1103/RevModPhys.81.109} { Rev. Mod. Phys. \textbf{81}, 109 (2019)}.
\bibitem{Wang}  W.-X. Wang, L.-J. Yin, J.-B. Qiao, T. Cai, S.-Y. Li, R.-F. Dou, J.-C. Nie, X. Wu, and L. He, Atomic resolution imaging of the two-component Dirac-Landau levels in a gapped graphene monolayer, \href{https://doi.org/10.1103/PhysRevB.92.165420}{Phys. Rev. B \textbf{92} ,165420 (2015)}.
\bibitem{Pilkyung1}  Pilkyung Moon and Mikito Koshino, Optical properties of the Hofstadter butterfly in the moir\'{e} superlattice, \href{https://journals.aps.org/prb/abstract/10.1103/PhysRevB.88.241412}{Phys. Rev. B \textbf{88} ,241412(R) (2013)}.
\bibitem{Pilkyung2}  J. A. Crosse, Pilkyung Moon, Faraday rotations, ellipticity and circular dichroism in the magneto-optical spectrum of moir\'{e} superlattices, \href{https://iopscience.iop.org/article/10.1088/1674-1056/ac051f}{Chin. Phys. B \textbf{30} ,077803 (2021)}.
\bibitem{Naumis2021} Leonardo A. Navarro-Labastida, Enrique Aguilar-Méndez, and Abdiel Espinosa-Champo, Atomic resolution imaging of the two-component Dirac-Landau levels in a gapped graphene monolayer, \href{https://journals.aps.org/prb/abstract/10.1103/PhysRevB.103.245418}{Phys. Rev. B \textbf{103} ,245418 (2021)}.
\bibitem{Arbeitman1}  J. Herzog-Arbeitman, A. Chew, D. K. Efetov and B. A. Bernevig, Reentrant correlated insulators in twisted bilayer graphene at 25T (2$\pi$ flux), \href{https://arxiv.org/abs/2111.11434}{arXiv:2111.11434}.
\bibitem{Arbeitman2}  I. Das, C. Shen, A. Jaoui, J. Herzog-Arbeitman, A. Chew, C.-W. Cho, T. Taniguchi, K. Watanabe, B. Piot, B. A. Bernevig and D. K. Efetov, Observation of re-entrant correlated insulators and interaction driven Fermi surface reconstructions at one magnetic flux quantum per moir\'{e} unit cell in magic-angle twisted bilayer graphene, \href{https://arxiv.org/abs/2111.11341}{arXiv:2111.11341}.

\bibitem{andrews_Fractional_2020} Bartholomew Andrews, Alexey Soluyanov, Fractional quantum Hall states for moir\'{e} superstructures in the Hofstadter regime, \href{https://journals.aps.org/prb/abstract/10.1103/PhysRevB.101.235312} {Phys. Rev. B \textbf{101}, 235312 (2020)}.

\bibitem{San} P. San-Jose, A. Gutiérrez-Rubio, M. Sturla, and F.
Guinea, Spontaneous strains and gap in graphene on boron nitride, \href{https://doi.org/10.1103/PhysRevB.90.075428} {Phys. Rev. B \textbf{90}, 075428 (2014)}.
\bibitem{Kindermann} M. Kindermann, B. Uchoa, and D. L. Miller, Zero-energy modes and gate-tunable gap in graphene on hexagonal boron nitride, \href{https://doi.org/10.1103/PhysRevB.86.115415} {Phys. Rev.
B \textbf{86}, 115415 (2012)}.
\bibitem{Zarenia} M. Zarenia, O. Leenaerts, B. Partoens, and F. M. Peeters, Substrate-induced chiral states in graphene, \href{https://doi.org/10.1103/PhysRevB.86.085451}{Phys. Rev. B \textbf{86}, 085451 (2012)}.
\bibitem{Zhou} S.Y. Zhou, D. A. Siegel, A. V. Fedorov, A. Lanzara, Metal to Insulator Transition in Epitaxial Graphene Induced by Molecular Doping, \href{https://doi.org/10.1103/PhysRevLett.101.086402} {Phys.
Rev. Lett. \textbf{101}, 086402 (2008)}.
\bibitem{Costa}  R. N. Costa Filho, G. A. Farias, and F. M. Peeters, Graphene ribbons with a line of impurities: Opening of a gap, \href{https://doi.org/10.1103/PhysRevB.76.193409}{Phys. Rev. B \textbf{76}, 193409 (2007)}.
\bibitem{Ben} B. Van Duppen, and F. M. Peeters, Four-band tunneling in bilayer graphene, \href{http://dx.doi.org/10.1103/PhysRevB.87.205427}{Phys. Rev. B \textbf{87}, 205427 (2013)}.
\bibitem{Ben1} M. Van der Donck, F. M. Peeters, and B. Van Duppen, Transport properties of bilayer graphene in a strong in-plane magnetic field, \href{http://dx.doi.org/10.1103/PhysRevB.93.115423}{Phys. Rev. B \textbf{93}, 115423 (2016)}.
\bibitem{Barbier} M. Barbier, P. Vasilopoulos, and F. M. Peeters, Kronig-Penney model on bilayer graphene: Spectrum and transmission periodic in the strength of the barriers, \href{https://doi.org/10.1103/PhysRevB.82.235408}{Phys. Rev. B \textbf{82}, 235408}.
\bibitem{Snyman} I. Snyman and C. W. J. Beenakker, Ballistic transmission through a graphene bilayer, \href{http://dx.doi.org/10.1103/PhysRevB.75.045322}{ Phys. Rev. B \textbf{75}, 045322 (2007)}.

\bibitem{MacCann} E. McCann and V. I. Fal’ko, Landau-Level Degeneracy and Quantum Hall Effect in a Graphite Bilayer,  \href{https://doi.org/10.1103/PhysRevLett.96.086805} {Phys. Rev. Lett. \textbf{96}, 086805 (2006)}.
\bibitem{Rozhkov} A. Rozhkov, A. Sboychakov, A. Rakhmanov, and F. Nori, Electronic properties of graphene-based bilayer systems, \href{https://doi.org/10.1016/j.physrep.2016.07.003} {Phys. Rep. \textbf{648}, 1 (2016)}
\bibitem{Ohta} T. Ohta, A. Bostwick, T. Seyller, K. Horn, and E. Rotenberg, Controlling the Electronic Structure of Bilayer Graphene, \href{https://doi.org/10.1126/science.1130681}{ Science \textbf{313}, 951 (2016)}.
\bibitem{Benlakhouy} N. Benlakhouy, A. El Mouhafid, and A. Jellal, Transport properties in gapped bilayer graphene, \href{https://doi.org/10.1016/j.physe.2021.114835} {Physica E \textbf{134}, 114835 (2021)}.
\bibitem{Chiu} C.-W. Chiu, S.-C. Chen, Y.-C. Huang, F.-L. Shyu, and M.-F. Lin, Critical optical properties of AA-stacked multilayer graphenes, \href{https://doi.org/10.1063/1.4813112} {Appl. Phys. Lett. \textbf{103}, 041907 (2013)}.
\bibitem{Zahidi} Y. Zahidi, I. Redouani, and A. Jellal, Goos–H$\ddot{\text{a}}$nchen shifts in AA-stacked bilayer graphene superlattices, \href{https://doi.org/10.1016/j.physe.2016.03.022} {Physica E \textbf{71}, 259  (2016)}.


\bibitem{Navarro1} Naumis, Gerardo G. and Navarro-Labastida, Leonardo A. and Aguilar-M\'endez, Enrique and Espinosa-Champo, Abdiel, Reduction of the twisted bilayer graphene chiral Hamiltonian into a $2\ifmmode\times\else\texttimes\fi{}2$ matrix operator and physical origin of flat bands at magic angles,  \href{https://doi.org/10.1103/PhysRevB.103.245418} {Phys. Rev. B \textbf{103}, 0245418 (2021)}.

\bibitem{Navarro2} Navarro-Labastida, Leonardo A. and Espinosa-Champo, Abdiel and Aguilar-Mendez, Enrique and Naumis, Gerardo G., Why the first magic-angle is different from others in twisted graphene bilayers: Interlayer currents, kinetic and confinement energy, and wave-function localization,  \href{https://doi.org/10.1103/PhysRevB.105.115434} {Phys. Rev. B \textbf{105}, 115434 (2021)}.


\bibitem{Cao} Y. Cao, V. Fatemi, S. Fang, K. Watanabe, T. Taniguchi, E. Kaxiras, and P. Jarillo-Herrero, Unconventional superconductivity in magic-angle graphene superlattices, \href{https://doi.org/10.1038/nature26160} {Nature \textbf{556}, 43 (2018)}.

\bibitem{Culchac}   F. J. Culchac, R. R. Del Grande, R. B. Capaz, L. Chico, and E. S. Morell, Flat bands and gaps in twisted double bilayer graphene, \href{https://doi.org/10.1039/C9NR10830K} {Nanoscale \textbf{12}, 5014 (2020)}.
\bibitem{Cao1} Y. Cao, D. Rodan-Legrain, O. Rubies-Bigorda, J. M. Park, K. Watanabe, T. Taniguchi, and P. Jarillo-Herrero, Tunable correlated states and spin-polarized phases in twisted bilayer–bilayer graphene, 
\href{https://doi.org/10.1038/s41586-020-2260-6} {Nature \textbf{583}, 215–220 (2020)}.
\bibitem{Shen} C. Shen, Y. Chu, Q. Wu, N. Li, S. Wang, Y. Zhao,
J. Tang, J. Liu, J. Tian, K. Watanabe, T. Taniguchi,
R. Yang, Z. Y. Meng, D. Shi, O. V. Yazyev, and
G. Zhang, Correlated states in twisted double bilayer
graphene, \href{https://doi.org/10.1038/s41567-020-0825-9} {Nature Physics \textbf{16}, 520 (2020)}.
\bibitem{Liu} X. Liu, Z. Hao, E. Khalaf, J. Y. Lee, Y. Ronen, H. Yoo, D. Haei Najafabadi, K. Watanabe, T. Taniguchi, A. Vish-wanath, and P. Kim, Tunable spin-polarized correlated
states in twisted double bilayer graphene, \href{https://doi.org/10.1038/s41586-020-2458-7} {Nature \textbf{583}, 221 (2020)}.

\bibitem{Lee} J. Y. Lee, E. Khalaf, S. Liu, X. Liu, Z. Hao, P. Kim, and A. Vishwanath, Theory of correlated insulating behaviour and spin-triplet superconductivity in twisted
double bilayer graphene, \href{https://doi.org/10.1038/s41467-019-12981-1} {Nature Communications \textbf{10}, 5333 (2019)}.

\bibitem{Kerelsky} A. Kerelsky, C. Rubio-Verd\'{u}, L. Xian, D. M. Kennes, D. Halbertal, N. Finney, L. Song, S. Turkel, L. Wang,
K. Watanabe, T. Taniguchi, J. Hone, C. Dean, D. Basov,
A. Rubio, and A. N. Pasupathy, Moir\'{e}-less correlations in
abca graphene, \href{https://doi.org/10.1073/pnas.2017366118}{10.1073/pnas.2017366118 (2021)}.

\bibitem{Rubio} C. Rubio-Verd\'{u}, S. Turkel, L. Song, L. Klebl, R. Samaj-dar, M. S. Scheurer, J. W. F. Venderbos, K. Watanabe, T. Taniguchi, H. Ochoa, L. Xian, D. Kennes, R. M. Fer-nandes, Angel Rubio, and A. N. Pasupathy, Universal 
moir\'{e} nematic phase in twisted graphitic systems,
\href{https://arxiv.org/abs/2009.11645} {arXiv:2009.11645}.
\bibitem{Halbertal} D. Halbertal, N. R. Finney, S. S. Sunku, A. Kerelsky, C. Rubio-Verd\'{u}, S. Shabani, L. Xian, S. Carr, S. Chen, C. Zhang, L. Wang, D. Gonzalez-Acevedo, A. S. McLeod,
D. Rhodes, K. Watanabe, T. Taniguchi, E. Kaxiras, C. R.
Dean, J. C. Hone, A. N. Pasupathy, D. M. Kennes, A. Rubio, and D. N. Basov, Moir\'{e} metrology of energy land-scapes in van der Waals heterostructures, \href{https://doi.org/10.1038/s41467-020-20428-1} {Nature Communications \textbf{12}, 242 (2021)}.
\bibitem{Christos} M. Christos, S. Sachdev, and M. S. Scheurer, Correlated insulators, semimetals, and superconductivity in
twisted trilayer graphene (2021), \href{https://arxiv.org/abs/2106.02063} {arXiv:2106.02063}.
\bibitem{Wong} D. Wong, K. P. Nuckolls, M. Oh, B. Lian, Y. Xie, S. Jeon, K. Watanabe, T. Taniguchi, B. A. Bernevig, and A. Yaz-dani, Cascade of electronic transitions in magic-angle
twisted bilayer graphene, \href{https://doi.org/10.1038/s41586-020-2339-0} {Nature \textbf{582}, 198 (2020)}.
\bibitem{Lu} X. Lu, P. Stepanov, W. Yang, M. Xie, M. A. Aamir,
I. Das, C. Urgell, K. Watanabe, T. Taniguchi, G. Zhang,
A. Bachtold, A. H. MacDonald, and D. K. Efetov, Superconductors, orbital magnets and correlated states in magic-angle bilayer graphene, \href{https://doi.org/10.1038/s41586-019-1695-0} {Nature \textbf{574}, 653 (2019)}.
\bibitem{Sharpe} A. L. Sharpe, E. J. Fox, A. W. Barnard,
J. Finney, K. Watanabe, T. Taniguchi, M. A.
Kastner, and D. Goldhaber-Gordon, Emergent
ferromagnetism near three-quarters filling in
twisted bilayer graphene, \href{https://doi.org/10.1126/science.aaw3780} {Science \textbf{365}, 605 (2019)}.
\bibitem{Seo} K. Seo, V. N. Kotov, and B. Uchoa, Ferromagnetic
mott state in twisted graphene bilayers at the magic
angle, \href{https://doi.org/10.1103/PhysRevLett.122.246402} { Phys. Rev. Lett. \textbf{122}, 246402 (2019)}.
\bibitem{Hass} J. Hass, F. Varchon, J. E. Millan-Otoya, M. Sprinkle, N. Sharma, W. A. deHeer, C. Berger, P. N. First, L. Magaud, and E. H. Conrad, Why Multilayer Graphene on 4H-SiC(0001¯) Behaves Like a Single Sheet of Graphene, \href{https://doi.org/10.1103/PhysRevLett.100.125504}{
Phys. Rev. Lett. \textbf{100}, 125504 (2008)}.
\bibitem{Brihuega} I. Brihuega, P. Mallet, H. Gonz$\acute{\text{a}}$lez-Herrero, G. Trambly de Laissardi\`{e}re, M. M. Ugeda, L. Magaud, J. M. G\'{o}mez-Rodr\'{\i}guez, F. Yndur$\acute{\text{a}}$in, and J.-Y. Veuillen, Unraveling the Intrinsic and Robust Nature of van Hove Singularities in
Twisted Bilayer Graphene by Scanning Tunneling Microscopy and Theoretical Analysis, \href{https://doi.org/10.1103/PhysRevLett.109.196802}{Phys. Rev. Lett. \textbf{109}, 196802
(2012)}.
\bibitem{Meng} L. Meng, Y. Zhang, W. Yan, L. Feng, L. He, R.-F. Dou, and J.-C. Nie, Single-layer behavior and slow carrier density dynamic of twisted graphene bilayer, \href{ https://doi.org/10.1063/1.4839419} {Appl. Phys. Lett. \textbf{100}, 091601 (2012)}.
\bibitem{Luican1} A. Luican, G. Li, A. Reina, J. Kong, R. R. Nair, K. S. Novoselov, A. K. Geim, and E. Y. Andrei, Single-Layer Behavior and Its Breakdown in Twisted Graphene Layers, \href{https://doi.org/10.1103/PhysRevLett.106.126802} {Phys. Rev. Lett. \textbf{106}, 126802 (2011)}.
\bibitem{Yan} W. Yan, M. Liu, R.-F. Dou, L. Meng, L. Feng, Z.-D. Chu, Y. Zhang, Z. Liu, J.-C. Nie, and L. He, Angle-Dependent van Hove Singularities in a Slightly Twisted Graphene Bilayer, \href{https://doi.org/10.1103/PhysRevLett.109.126801} {Phys. Rev. Lett. \textbf{109}, 126801 (2012)}.
\bibitem{Li1} G. Li, A. Luican, J. M. B. Lopes dos Santos, A. H. Castro Neto, A. Reina, J. Kong, and E. Y. Andrei, Observation of Van Hove singularities in twisted graphene layers, \href{https://doi.org/10.1038/nphys1463} {Nat. Phys. \textbf{6}, 109 (2010)}.
\bibitem{Cisternas} E. Cisternas and J. Correa, Theoretical reproduction of superstructures revealed by STM on bilayer graphene, \href{https://doi.org/10.1016/j.chemphys.2012.09.021} {Phys. \textbf{409}, 74 (2012)}.

\bibitem{Oka} T. Oka and S. Kitamura, Floquet engineering of quantum materials, \href{https://doi.org/10.1146/annurev-conmatphys-031218-013423} {Annual Review of Condensed Matter Physics
\textbf{10}, 387–408 (2019)}.
\bibitem{Oka1} T. Oka and H. Aoki, Photovoltaic hall effect in graphene, Phys. Rev. B \textbf{79}, 081406 (2009).
\bibitem{Luo} M. Luo, Tuning of a bilayer graphene heterostructure by horizontally incident circular polarized light, \href{https://doi.org/10.1103/PhysRevB.103.195422} {Phys. Rev. B \textbf{103}, 195422 (2021)}.
\bibitem{Kibis} O. V. Kibis, S. Morina, K. Dini, and I. A. Shelykh, Magnetoelectronic properties of graphene dressed by a high-frequency field, \href{https://doi.org/10.1103/PhysRevB.93.115420} {Phys. Rev. B \textbf{93}, 115420 (2016)}.

\bibitem{Vogl4}	 M. Rodriguez-Vega, M. Vogl, G. A. Fiete, Low-frequency and Moir\'{e} Floquet engineering: Areview, \href{https://doi.org/10.1016/j.aop.2021.168434} {Annals of Physics, 168434 (2021)}.

\bibitem{Albanin} D. A. Abanin, W. De Roeck, W. W. Ho, and F. m. c. Huveneers, Effective Hamiltonians, prethermalization, and
slow energy absorption in periodically driven many-body
systems, \href{https://doi.org/10.1103/PhysRevB.95.014112} {Phys. Rev. B \textbf{95}, 014112 (2017)}.
\bibitem{Itin} A. P. Itin and M. I. Katsnelson, Effective Hamiltonians for rapidly driven many-body lattice systems: Induced exchange interactions and density-dependent hoppings, \href{https://doi.org/10.1103/PhysRevLett.115.075301} {
Phys. Rev. Lett. \textbf{115}, 075301 (2015)}.
\bibitem{Mikami} T. Mikami, S. Kitamura, K. Yasuda, N. Tsuji, T. Oka, and H. Aoki, Brillouin-Wigner theory for high-frequency
expansion in periodically driven systems: Application to
Floquet topological insulators, \href{https://doi.org/10.1103/PhysRevB.93.144307} { Phys. Rev. B \textbf{93}, 144307
(2016)}.
\bibitem{Mohan} P. Mohan, R. Saxena, A. Kundu, and S. Rao, Brillouin-Wigner theory for Floquet topological phase transitions
in spin-orbit-coupled materials, \href{https://doi.org/10.1103/PhysRevB.94.235419} { Phys. Rev. B \textbf{94}, 235419
(2016)}.
\bibitem{Vogl5} M. Vogl, P. Laurell, A. D. Barr, and G. A. Fiete, Analog of hamilton-jacobi theory for the time-evolution operator, \href{https://doi.org/10.1103/physreva.100.012132} {Physical Review A \textbf{100}, (2019)}.
\bibitem{Vogl1} M. Vogl, M. Rodriguez-Vega, G. A. Fiete, Effective Floquet Hamiltonians for periodically driven twisted bilayer graphene, \href{https://doi.org/10.1103/PhysRevB.101.235411} {Phys Rev B \textbf{101}, 235411 (2020)}.
\bibitem{Vogl6} M. Vogl, P. Laurell, A. D. Barr, and G. A. Fiete, Flow equation approach to periodically driven quantum systems, \href{https://doi.org/10.1103/PhysRevX.9.021037} {Phys. Rev. X \textbf{9}, 021037 (2019)}.
 
\bibitem{Vega} M. Rodriguez-Vega, M. Lentz, and B. Seradjeh, Floquet perturbation theory: formalism and application to
low-frequency limit, \href{https://doi.org/10.1088/1367-2630/aade37} {New Journal of Physics \textbf{20}, 093022
(2018)}.
\bibitem{Vega1} M. Rodriguez-Vega and B. Seradjeh, Universal fluctuations of floquet topological invariants at low frequencies, \href{https://doi.org/10.1103/PhysRevLett.121.036402} {
Phys. Rev. Lett. \textbf{121}, 036402 (2018)}.

\bibitem{Kennes} D. M. Kennes, N. M\"{u}ller, M. Pletyukhov, C. Weber, C. Bruder, F. Hassler, J. Klinovaja, D. Loss, and
H. Schoeller, Chiral one-dimensional floquet topological insulators beyond the rotating wave approximation, \href{https://doi.org/10.1103/PhysRevB.100.041103} {Phys. Rev. B \textbf{100}, 041103 (2019)}.
\bibitem{Li11} Z.-Z. Li, C.-H. Lam, and J. Q. You, Floquet engineering of long-range p-wave superconductivity: Beyond the
high-frequency limit, \href{ https://doi.org/10.1103/PhysRevB.96.155438} {Phys. Rev. B \textbf{96}, 155438 (2017)}.
\bibitem{Muller} N. M\"{u}ller, D. M. Kennes, J. Klinovaja, D. Loss, and H. Schoeller, Electronic transport in one-dimensional floquet topological insulators via topological and nontopological edge states, \href{https://doi.org/10.1103/PhysRevB.101.155417} {Phys. Rev. B \textbf{101}, 155417 (2020)}.

\bibitem{Ibsal} I. A. Assi, J. P. F. LeBlanc, M. Rodriguez-Vega, H. Bahlouli, and M. Vogl, Floquet engineering and non-equilibrium topological maps in twisted trilayer graphene, \href{https://doi.org/10.1103/PhysRevB.104.195429}{Phys. Rev. B \textbf{104}, 195429 (2021)}.

\bibitem{Topp} G. E. Topp, G. Jotzu, J. W. McIver, L. Xian, A. Rubio, and M. A. Sentef, Topological floquet engineering of twisted bilayer graphene, \href{https://doi.org/10.1103/physrevresearch.1.023031} {Physical Review Research \textbf{1}, 10.1103 (2019)}.
\bibitem{Vogl2} M. Vogl, M. Rodriguez-Vega, G. A. Fiete, Floquet engineering of interlayer couplings: Tuning the magic angle of twisted bilayer graphene at
the exit of a waveguide, \href{https://doi.org/10.1103/PhysRevB.101.241408} { Phys Rev B \textbf{101}, 241408(R) (2020)}.
\bibitem{Lu1} M. Lu, J. Zeng, H. Liu, J.-H. Gao, and X. C. Xie, Valley-selective floquet chern flat bands in twisted multilayer
graphene, \href{https://doi.org/10.1103/PhysRevB.103.195146} {Phys. Rev. B \textbf{103}, 195146 (2021)}.
\bibitem{Vega2} M. Rodriguez-Vega, M. Vogl, and G. A. Fiete, Floquet
engineering of twisted double bilayer graphene, \href{https://doi.org/10.1103/PhysRevResearch.2.033494} {Phys.
	Rev. Research \textbf{2}, 033494 (2020)}.

\bibitem{MacDonlad} R. Bistritzer and A. H. MacDonald, moir\'{e} butterflies in twisted bilayer graphene, \href{https://doi.org/10.1103/PhysRevB.84.035440} {Phys. Rev B \textbf{84}, 035440 (2011)}.


\bibitem{Hofstadter} D. R. Hofstadter, Energy levels and wave functions of Bloch electrons in rational and irrational magnetic fields, \href{https://doi.org/10.1103/PhysRevB.14.2239} { Phys. Rev. B \textbf{14}, 2239 (1976)}.
\bibitem{Hunt} B. Hunt, J. D. Sanchez-Yamagishi, A. F. Young, M. Yankowitz, B. J. LeRoy, K. Watanabe, T. Taniguchi, P. Moon,
M. Koshino, P. Jarillo-Herrero, and R. C. Ashoori, Massive Dirac Fermions and Hofstadter Butterfly in a van der Waals Heterostructure, \href{https://doi.org/10.1126/science.1237240} {Science \textbf{340}, 1427 (2013)}.

\bibitem{Ponomarenko} L. A. Ponomarenko, R. V. Gorbachev, G. L. Yu, D. C. Elias, R. Jalil, A. A. Patel, A. Mishchenko, A. S. Mayorov, C. R. Woods, J. R. Wallbank, M. Mucha-Kruczynski, B. A. Piot, M. Potemski, I. V. Grigorieva, K. S. Novoselov, F. Guinea, V. I. Fal’ko, and A. K. Geim, Cloning of Dirac fermions in graphene superlattices, \href{ https://doi.org/10.1038/nature12187} {Nature \textbf{497}, 594 (2013)}.
\bibitem{Dean} C. R. Dean, L. Wang, P. Maher, C. Forsythe, F. Ghahari, Y. Gao, J. Katoch, M. Ishigami, P. Moon, M. Koshino, et al, Hofstadter’s butterfly and the fractal quantum Hall effect in moir\'{e} superlattices, \href{https://doi.org/10.1038/nature12186} {Nature \textbf{497}, 598 (2013)}.
\bibitem{Oh} G.-y. Oh, Energy Spectrum of a Triangular Lattice in a Uniform Magnetic Field: Effect of Next-Nearest-Neighbor Hopping, \href{https://doi.org/10.3938/jkps.37.534} {J. Korean Phys. Soc. \textbf{37}, 534 (2000)}.
\bibitem{Oh1} G.-Y. Oh, Energy Spectrum of a Honeycomb Lattice under Nonuniform
Magnetic Fields, J. Korean Phys. Soc. \textbf{49}, 672 (2006).
\bibitem{Du} L. Du, Q. Chen, A. D. Barr, A. R. Barr, and G. A. Fiete, Floquet Hofstadter butterfly on the kagome and triangular lattices, \href{https://doi.org/10.1103/PhysRevB.98.245145} {Phys. Rev B \textbf{98}, 245145 (2018)}.

\bibitem{Bistritzer} R. Bistritzer and A. H. MacDonald, moir\'{e} bands in twisted double-layer graphene, \href{ https://doi.org/10.1073/pnas.1108174108 } {Proc. Natl. Acad. Sci. \textbf{108}, 12233 (2011)}.



\bibitem{Kasra} K. Hejazi, C. Liu, and L. Balents, Landau levels in twisted bilayer graphene and semiclassical orbits, \href{https://doi.org/10.1103/PhysRevB.100.035115}{Phys. Rev. B \textbf{100}, 035115 (2019)}
\bibitem{Kasra2} Kasra Hejazi, Chunxiao Liu, Hassan Shapourian, Xiao Chen, and Leon Balents, Multiple topological transitions in twisted bilayer graphene near the first magic angle, \href{https://journals.aps.org/prb/abstract/10.1103/PhysRevB.99.035111}{Phys. Rev. B \textbf{99}, 035111 (2019)}
\bibitem{Zhang1} Y. H Zhang , H. C. Po, and T. Senthil, Landau level degeneracy in twisted bilayer graphene: Role of symmetry breaking, \href{https://doi.org/10.1103/PhysRevB.100.125104} {Phys. Rev. B \textbf{100}, 125104 (2019)}.

\bibitem{Wang2} J. Wang, A. S. Mouritzen, and J. Gong, Quantum control of ultra-cold atoms: uncovering a novel connection between two paradigms of quantum nonlinear dynamics, \href{https://doi.org/10.1080/09500340802187365} { J. Mod. Opt. \textbf{56}, 722 (2009)}.
\bibitem{Lawton} W. Lawton, A. S. Mouritzen, Spectral relationships between kicked Harper and on-resonance double kicked rotor operators, \href{https://doi.org/10.1063/1.3085756}{
	Phys. \textbf{50}, 032103 (2009)}.
\bibitem{Wang3}H. Wang, D. Y. H. Ho, W. Lawton, J. Wang, and J. Gong, Kicked-Harper model versus on-resonance double-kicked rotor model: From spectral difference to topological equivalence, \href{https://doi.org/10.1103/PhysRevE.88.052920} {
	Phys. Rev. E \textbf{88}, 052920 (2013)}.
\bibitem{Labadidi} M. Lababidi, I. I. Satija, and E. Zhao, Counter-propagating Edge Modes and Topological Phases of a Kicked Quantum Hall System, \href{https://doi.org/10.1103/PhysRevLett.112.026805} { Phys. Rev. Lett. \textbf{112},
	026805 (2014)}.
\bibitem{Zhou1} Z. Zhou, I. I. Satija, and E. Zhao, Floquet edge states in a harmonically driven integer quantum Hall system, \href{https://doi.org/10.1103/PhysRevB.90.205108} { Phys. Rev. B \textbf{90}, 205108
	(2014)}.
\bibitem{Ding} K.-H. Ding, L.-K. Lim, G. Su, and Z.-Y. Weng, Quantum Hall effect in ac driven graphene: From the half-integer to the integer case, \href{https://doi.org/10.1103/PhysRevB.97.035123} {Phys. Rev. B
	\textbf{97}, 035123 (2018)}.
\bibitem{Wackerl} M. Wackerl and J. Schliemann, Driven Hofstadter Butterflies and Related Topological Invariants, \href{https://doi.org/10.1103/PhysRevB.104.195429}{Phys. Rev. B \textbf{100}, 165411 (2019) }.
\bibitem{Kooi}S. H. Kooi, A. Quelle, W. Beugeling, and C. Morais Smith, Genesis of the Floquet Hofstadter butterfly, \href{https://doi.org/10.1103/PhysRevB.98.115124} {
	Phys. Rev. B \textbf{98}, 115124 (2018)}.
\bibitem{JDJACKSON} J. D. Jackson, Classical electrodynamics, 3rd Edition, Wiley, New York, NY, 1999.
href{http://cdsweb.cern.ch/record/490457}


\bibitem{Wu} F. Wu, A. H. MacDonald, and I. Martin, Theory of Phonon-Mediated Superconductivity in Twisted Bilayer Graphene, \href{https://doi.org/10.1103/PhysRevLett.121.257001} {Phys. Rev. Lett. \textbf{121}, 257001 (2018)}.

\bibitem{Rost} F. Rost, R. Gupta, M. Fleischmann, D. Weckbecker, N. Ray, J. Olivares, M. Vogl, S. Sharma, O. Pankratov, and S. Shallcross, Nonperturbative theory of effective Hamiltonians for deformations in two-dimensional materials: moir\'{e} systems and
dislocations, \href{https://doi.org/10.1103/PhysRevB.100.035101} {Phys. Rev. B \textbf{100}, 035101 (2019)}.
\bibitem{Fleischmann} M. Fleischmann, R. Gupta, F. Wullschläger, S. Theil, D. Weckbecker, V. Meded, S. Sharma, B. Meyer, and S.
Shallcross, Perfect and controllable nesting in minimally twisted bilayer graphene, \href{https://doi.org/10.1021/acs.nanolett.9b04027} {Nano Lett. \textbf{20}, 971 (2020)}.
\bibitem{Xie} M. Xie and A. H. MacDonald, Nature of the Correlated Insulator States in Twisted Bilayer Graphen, \href{https://doi.org/10.1103/PhysRevLett.124.097601} {Phys. Rev. Lett. \textbf{124}, 097601 (2020)}.
\bibitem{Fleischmann1} M. Fleischmann, R. Gupta, S. Sharma, and S. Shallcross, Moir\'{e} quantum well states in tiny angle two dimensional semi-conductors, \href{https://arxiv.org/abs/1901.04679} {arXiv:1901.04679}.
\bibitem{Nam} N. N. T. Nam and M. Koshino, Lattice relaxation and energy band modulation in twisted bilayer graphene, \href{https://doi.org/10.1103/PhysRevB.96.075311} {Phys. Rev. B \textbf{96},
	075311 (2017)}.
\bibitem{Li} Y. Li, H. A. Fertig, and B. Seradjeh, Floquet-engineered topological flat bands in irradiated twisted bilayer graphene,  \href{https://doi.org/10.1103/PhysRevResearch.2.043275}{Phys. Rev. Research \textbf{2}, 043275 (2020)}.

\bibitem{Yang2020} Y. Yang, B. Zhen, J. D. Joannopoulos, and M. Soljačić, Non-Abelian generalizations of the Hofstadter
model: spin–orbit-coupled butterfly pairs, \href{https://doi.org/10.1038/s41377-020-00384-7}{Light. Sci. Appl \textbf{9}, 177 (2020)}.
\bibitem{carr} S. Carr, S. Fang, Z. Zhu, and E. Kaxiras, Exact continuum model for low-energy electronic states of twisted bilayer graphene, \href{https://doi.org/10.1103/PhysRevResearch.1.013001}{Phys. Rev. Research 1, 013001 (2019)}.




\bibitem{Dehghani} H. Dehghani, T. Oka, and A. Mitra, Out-of-equilibrium electrons and the Hall conductance of a Floquet topological insulator, \href{https://doi.org/10.1103/PhysRevB.91.155422}{Phys. Rev B \textbf{91}, 155422 (2015)}.

















 		
	\end{thebibliography}
\end{document}